\documentclass[twocolumn,prb,color,superscriptaddress,psfig]{revtex4}
\usepackage{graphicx}
\usepackage{epsfig}
\begin{document}
\title{Charge Transfer Excitations in   Insulating Copper Oxides}
\author{A.S. Moskvin}
\affiliation{Ural State University, 620083 Ekaterinburg,  Russia}
\author{S.-L. Drechsler}
\affiliation{Leibniz-Institut f\"{u}r Festk\"{o}rper- und
Werkstoffforschung Dresden, P.O.\ Box 270116, D-01171 Dresden,
Germany}
\author{R. Hayn}
\affiliation{Leibniz-Institut f\"{u}r Festk\"{o}rper- und
Werkstoffforschung Dresden, P.O.\ Box 270116, D-01171 Dresden,
Germany} \affiliation{Laboratoire Mat\'eriaux et Micro\'electronique
de Provence, Universit\'e Paul Cezanne, Case 142, F-13397 Marseille
Cedex 20, France}
\author{J. M\'{a}lek}
\affiliation{Leibniz-Institut f\"{u}r Festk\"{o}rper- und
Werkstoffforschung Dresden, P.O.\ Box 270116, D-01171 Dresden,
Germany} \affiliation{Institute of Physics, ASCR, Na Slovance 2,
CZ-18221 Praha 8, Czech Republic}
\date{\today}
\begin{abstract}
A semi-quantitative cluster approach is developed to describe the
charge transfer (CT) electron-hole excitations in insulating
cuprates in a rather wide energy range up to $10\div 15$ eV. It
generalizes the Zhang-Ng (ZN) model of CT excitons by considering
the complete set of Cu $3d$ and O $2p$ orbitals within the  CuO$_4$
embedded molecular  cluster method and by introducing one-center
(intra-center) Frenkel-like and two-center (inter-center) excitons.
Special attention is paid to the transition matrix element effects
both in optical and electron energy loss  spectra (EELS). In the
latter case we obtain the momentum dependence of matrix elements
both  for intra-center and inter-center transitions. We are able to
give a semi-quantitative description of the optical and EELS spectra
for a large number of 0D (like CuB$_2$O$_4$), 1D (Sr$_2$CuO$_3$) and
2D (like Sr$_2$CuO$_2$Cl$_2$) insulating cuprates in a unifying
manner. By comparing our analysis with the experimental data we find
that the CT gap in insulating cuprates is determined by nearly
degenerate intra-center localized excitations and inter-center CT
excitons. The former are associated with a hole CT transition
$b_{1g}\rightarrow e_{u}(\pi)$ from the $b_{1g}\propto d_{x^2-y^2}$
ground state of dominantly Cu $3d_{x^2-y^2}$ character to a purely
oxygen dominantly O $2p_\pi$ state localized on one CuO$_4$
plaquette, whereas the latter correspond to a $b_{1g}\rightarrow
b_{1g}$ CT transition between neighboring plaquettes with the
Zhang-Rice (ZR) singlet being the final two-hole state.  The
corresponding EELS intensity is found to be strongly ${\bf k}$
dependent: even for isolated exciton it sharply decreases by
approaching the Brillouin zone (BZ) boundary. It is shown that the
interaction of two-center excitons can result in destructive
interference effects with an intensity compensation point.
\end{abstract}

\maketitle

\section{Introduction}

The nature of the  electron-hole excitations in parent quasi-2D
cuprates such as La$_{2}$CuO$_{4}$, Sr$_2$CuO$_2$Cl$_2$,
YBa$_2$Cu$_3$O$_6$,  and their 1D counterparts like Sr$_2$CuO$_3$,
Sr$_2$CuO$_2$, Li$_2$CuO$_2$ represents an important challenging
issue both for the high-$T_c$ problem and, more generally, for
strongly correlated oxides. The mechanism of low-energy
electron-hole excitations as well as those with higher energy and
high intensity is still unclear.

It is now widely believed that  the most intensive low-energy
electronic excitations in quasi-2D insulating copper oxides
correspond to the transfer of electrons from O to Cu in the
CuO$_{2}$ layer, hence these materials are charge transfer (CT)
insulators.\cite{ZSA} Moreover, sometimes the intensive band near
$2\div 3$ eV  is naively supposed to be the only representative of
the O $2p$-Cu $3d$ charge transfer, and higher-lying structures are
assigned to transitions relevant to Cu $4s$, La $5d/4f$, Sr $5d$, or
Ca $4d$ depending on the actual chemical formula of the
corresponding cuprate.

Despite the giant ($\sim 10^4$) number of  experimental and
theoretical papers on cuprates we deal actually with a lack of
detailed studies of electron-hole excitations both in parent and
high-$T_c$ cuprates. The most part of optical information
\cite{Tokura} was  obtained to date by reflectivity measurements
followed by a Kramers-Kronig transformation usually accompanied by a
number of unavoidable uncertainties as regards the peak positions
and intensities of weak spectral features. Nevertheless, applying
different optical  measurements including Raman spectroscopy and
nonlinear susceptibility, one obtains important information, in
particular, as regards the subtleties of the optical gap states in
such insulating cuprates as CuO, \cite{Moskvin1994,Sukhorukov}
La$_2$CuO$_4$, \cite{Falck,Ohana,Kishida1,THG} R$_2$CuO$_4$ (R=La,
Nd, Eu, Gd), \cite{Krich,Kishida1} YBa$_2$Cu$_3$O$_6$,
\cite{Heyen,Cooper} Sr$_2$CuO$_2$Cl$_2$,
\cite{Choi,optics,Blumberg,Schumacher} and
Sr$_2$CuO$_3$.\cite{Ogasawara,nonlinear,Kishida}

Unfortunately, any conventional optical technique is momentum
restricted, and yields an $indirect$ information on the dielectric
function $\epsilon (\omega , {\bf k})$ only at the $\Gamma$-point
given by ${\bf k}\rightarrow 0$. In other words, such a technique
cannot probe the dynamics of electron-hole excitations.

In this connection, we would like to emphasize the decisive role of
$direct$ EELS measurements in the observation and analysis of
electron-hole excitations as compared with conventional $indirect$
optical data. The possibility to yield the polarization and momentum
dependent loss function $Im(-1/\epsilon (\omega , {\bf k}))$ makes
EELS a powerful tool to examine subtle details of the energy
spectrum and dynamics of electron-hole excitations in cuprates.
\cite{Wang,EELS,EELS1} In the limit ${\bf k}\rightarrow 0$ the
selection rules are the same as in optics, i.e. only dipole
transitions are allowed. For finite ${\bf k}$ nondipole transitions
are detected as well. In contrast to reflectivity measurements, the
EELS in transmission is a surface insensitive technique. Recently,
momentum resolved resonant inelastic X-ray scattering (RIXS)
measurements \cite{Abbamonte,Hasan} performed for
Sr$_2$CuO$_2$Cl$_2$ and  Ca$_2$CuO$_2$Cl$_2$  have demonstrated the
feasibility of RIXS to study electron-hole excitations in insulating
cuprates.

The theoretical analysis of optical excitations beyond simple band
models has been mostly performed in the frame of conventional
Hubbard-like models, where  one usually considers only O
$2p_{\sigma}$ orbitals, and neglects O $2p_{\pi}$ and O $2p_z$
orbitals at all. In the parameter regime appropriate for undoped
cuprates, one implies that the one hole per unit cell is mainly
localized on the Cu site. Application of the current operator  onto
this state provides a charge transfer from the Cu to the O site.
Thus, one would expect a dominant absorption feature at the energy
of the charge transfer gap. In terms of the Hubbard model, this is a
CT transition from the nonbonding oxygen to the upper Hubbard band.
It was emphasized that instead of a transition to the upper Hubbard
band, it is more reasonable to speak of a transition to the
correlated Zhang-Rice \cite{ZR} type states. \cite{Wagner}

One of the central issues in the analysis of electron-hole
excitations  is whether low-lying states   are comprised of free
charge carriers or excitons. A conventional approach implies that if
the Coulomb interaction is effectively screened and weak, then the
electrons and holes are only weakly bound and move essentially
independently as free charge-carriers. However, if the Coulomb
interaction between electron and hole is strong, excitons are
formed, i.e.\ bound particle-hole pairs with strong correlation of
their mutual motion. In practice, many authors consider excitons to
consist of real-space configurations with electrons and holes
occupying nearest neighbor sites, while the electrons and holes are
separated from each other in the conduction-band states.\cite{Guo}

The electron-hole excitations near the CT gap in insulating CuO$_2$
planes were theoretically examined in Ref.\ \onlinecite{Simon} in
terms of a six-band model and using the cell-perturbation method.
The excitons were approximately treated as electrons and holes
moving freely in their respective quasi-particle bands, except in
the nearest neighborhood, where they feel an attraction and an
on-site interaction determined by the eigenstate of the cell
Hamiltonian, which leads to a rather small exciton dispersion.

However, a small exciton dispersion is in contradiction to the
results of the EELS measurements for the 2D model cuprate
Sr$_2$CuO$_2$Cl$_2$. \cite{Wang} These measurements stimulated the
elaboration of a simple CT exciton model by Zhang and
Ng.\cite{Wang,Ng} They used a local model to study the formation and
the structure of the low-energy charge-transfer excitations in the
insulating CuO$_2$ plane restricting themselves to the Cu
$3d_{x^{2}-y^{2}}$ and O $2p_{\sigma}$ orbitals. The elementary
excitation is a bound exciton of spin singlet, consisting of a
Cu$^+$ and a neighboring Zhang-Rice singlet-like excitation of Cu-O
holes with a rather large dispersion. They considered four
eigenmodes of excitons with different symmetry $A_{1g}, B_{1g},
E_{u}$, or $S,D, P_{1,2}$. The experimental peak analysis in
Sr$_2$CuO$_2$Cl$_2$ revealed a surprisingly large energy dispersion
$\approx 1.5$ eV along $[110]$ which was interpreted in terms of a
small exciton moving through the lattice freely without disturbing
the antiferromagnetic spin background, in contrast to the single
hole motion.\cite{Wang,Ng} So, it seems that the situation in
antiferromagnetic cuprates differs substantially from that in usual
semiconductors or in other bandlike insulators where, as a rule, the
effective mass of the electron-hole pair is larger than that of an
unbound electron and hole.

Other theoretical works in the frame of conventional Hubbard-like
models, without the O $2p_\pi$ and $2p_z$ orbitals, are those by
Hanamura {\it et. al.} \cite{Hanamura} and Kuzian {\it et. al.}
\cite{Kuzian} The {\it excitonic cluster model} \cite{Hanamura} has
been elaborated to describe the CT transitions in insulating
cuprates in frame of the conventional three-band Hubbard
Hamiltonian. The authors made an attempt to treat both the bound and
unbound states of the CT electron-hole pair on the same footing. To
this end they performed perturbation calculation for the excited
states of $E_u$ symmetry generated by the oxygen O $2p_\sigma$ hole
separated from the Cu $3d_{x^2 -y^2}$ electron by distances of 1, 2,
3, 4, or 5 nearest neighbors. To evaluate the dipole transition
matrix element they took into account only the transition between
the $d_{x^2-y^2}$ and the O $2p_\sigma$ orbital of $e_u$ symmetry.
The frequency and wave number dependence of the dielectric function
$\epsilon ({\bf k},\omega)$ and its inverse $\epsilon ^{-1}({\bf
k},\omega)$ has also been studied in Ref.\ \onlinecite{Kuzian}. The
authors show that the problem, in general, cannot be reduced to a
calculation within the single band Hubbard model, which takes into
account only a restricted number of electronic states near the Fermi
energy. The contribution of the rest of the system to the
longitudinal response is essential for the whole frequency range.
With the use of the spectral representation of the two-particle
Green's function they show that the problem may be divided into two
parts: into the contributions of the weakly correlated and the
Hubbard subsystems. For the latter  an approach is proposed that
starts from the correlated paramagnetic ground state with strong
antiferromagnetic fluctuations. The method is applied to the
multiband Hubbard (Emery) model that describes layered cuprates.

Already a shorthand inspection of the original experimental
data,\cite{Wang} and especially of the high-resolution EELS spectra
\cite{EELS,EELS1} for Sr$_2$CuO$_2$Cl$_2$ point to some essential
shortcomings of the Zhang-Ng (ZN) model. The high-resolution EELS
spectra \cite{EELS,EELS1} for insulating Sr$_2$CuO$_2$Cl$_2$ with a
momentum range along [110] broader than in Ref.\ \onlinecite{Wang}
did not confirm two principal predictions of the ZN-model: no spots
of the predicted $D$-mode exciton were found in a broad energy range
below $2$ eV; a clear decrease contrary to the predicted increase of
EELS intensity along $[100]$ was observed with momentum  increase
from $k=0.5$ to $k=0.7$. Moreover, the high-resolution measurements
\cite{EELS,EELS1} unambiguously detected a "multi-excitonic" nature
of EELS spectra in the $2\div 8$ eV energy range with a clear
manifestation of several dispersionless narrow bands with energies
near $2.0, 4.2, 5.4$, and $7.2$ eV. Thus, we conclude that the
simple ZN model should be revisited to consistently explain  the
large body of new experimental data.

Especially important is the proper inclusion of the O $2p_\pi$
orbitals into the theory of electron-hole excitations. One of the
first microscopic models for the low-energy CT transition in the
insulating copper oxide CuO peaked near 1.7 eV was given in Ref.\
\onlinecite{Moskvin1994}. This transition was assigned to the
dipole-allowed CT transition $b_{1g}\rightarrow e_u$ from the Cu
$3d$-O $2p$ hybrid $b_{1g}\propto d_{x^2 -y^2}$ hole ground state to
the purely oxygen orbital doublet $e_u$ state with predominant O
$2p_\pi$ weight. To the best of our knowledge, it was the first
indication of an optical manifestation of low-lying nonbonding O
$2p_\pi$ states.

In the present work we generalize the ZN-theory into a simple,
physically clear, cluster theory for the CT excitons that on the one
hand catches the essential physics and on the other hand provides a
semi-quantitative description of the optical and EELS spectra. A
preliminary analysis \cite{PRB,PRL} has shown that a concise
interpretation of the experimental EELS spectra may be obtained by
developing further the ZN-model with respect to two points. First of
all, we include the complete set of Cu $3d$ and O $2p$ orbitals of
the  plaquette in terms of the embedded cluster method.  And second,
we argue that the elementary O-Cu CT process \cite{Wang,Ng}
generates two types of excitons: one-center excitons localized on
one  CuO$_4$ plaquette and two-center excitons extending over two
CuO$_4$ plaquettes. We base our argumentation mainly on the ${\bf
k}$ dependence of the corresponding EELS intensity as revealed by a
detailed analysis of matrix elements of one- and two-center
excitons. We also investigate the exciton dynamics. Finally we
arrive at a semi-quantitative understanding of the electronic states
that dominate the optical and EELS response in a rather broad energy
range. In such a way we explain the main features of optical and
angle-resolved EELS spectra for a large number of different
cuprates. They reach from the 0D compound CuB$_2$O$_4$, over the 1D
cuprate Sr$_2$CuO$_4$ up to the 2D model compound
Sr$_2$CuO$_2$Cl$_2$ and the parent cuprates of high-$T_c$
superconductors. For instance, we present an alternative explanation
of the angle-resolved EELS spectra of Sr$_2$CuO$_2$Cl$_2$ that is
free from many shortcomings of the ZN-model.
We should emphasize that our theory, like also those of Zhang and Ng
cannot provide a proof whether the low-lying particle-hole pairs are
bound or not. It has to be expected that at least the higher lying
particle-hole excitations on a finite cluster do in reality not
exist as quasi-particle excitations with infinite life time.
Nevertheless, they should be visible in the spectra as resonances,
having possibly a rather large broadening. In that sense, we will
denote $all$ particle-hole excitations as $excitons$ in the present
paper. The question of its life time has to be answered
independently. Therefore, we cannot contribute for example to the
ongoing discussion about the stability of long-lived excitons in 1D
cuprates. \cite{Penc,Tsutsui,Kim04}

The rest of the paper is organized as follows. In Sec.\ II we
shortly address the electronic structure and the energy spectrum of
a CuO$_4$ cluster. The one-center small excitons with appropriate
transition matrix elements are considered in Sec.\ III. The
two-center small excitons and accompanying issues of correlations,
final state effects, EELS transition matrix elements, even and odd
excitons are addressed in Sec.\ IV. In Sec.\ V we  address the
problem of the energy and intensity  dispersion of two-center
excitons. Sec.\ VI is devoted to a model analysis of the
experimental EELS spectra in 0D, 1D and 2D insulating cuprates.

\section{Electronic structure of copper-oxygen clusters}

The electronic states in  strongly correlated cuprates manifest
both significant correlations and dispersional features. The
dilemma posed by such a combination is the overwhelming number of
configurations which must be considered in treating strong
correlations in a truly bulk system. One strategy to deal with
this dilemma is to restrict oneself to small clusters, creating
model Hamiltonians whose spectra may reasonably well represent the
energy and dispersion of the important excitations of the full
problem. Naturally, such an approach has a number of principal
shortcomings, including the boundary conditions, the breaking of
local symmetry of boundary atoms, and so on.

As an efficient approach to describe excitonic states, especially
with small effective electron-hole separation we propose here the
embedded molecular cluster method. In the present context we use
one or two neighboring CuO$_4$ clusters embedded into the
insulating cuprate. This method provides both, a clear physical
picture of the complex electronic structure and the energy
spectrum, as well as the possibility of quantitative modelling.
Eskes {\it et al.} \cite{Eskes}, as well as Ghijsen {\it et al.}
\cite{Ghijsen} have shown that in a certain sense the cluster
calculations might provide a better description of the overall
electronic structure of  insulating  copper oxides  than
band-structure calculations.  They allow to take better into
account correlation effects.

Beginning from 5 Cu $3d$ and 12 O $2p$ atomic orbitals for CuO$_4$
 cluster with $D_{4h}$ symmetry, it is easy to form 17 symmetrized
 $a_{1g},a_{2g},b_{1g},b_{2g},e_{g}$ (gerade=even) and $a_{2u},b_{2u},
 e_{u}(\sigma),e_{u}(\pi)$
 (ungerade=odd) orbitals. The even Cu $3d$ $a_{1g}(3d_{z^2}), b_{1g}(3d_{x^2-y^2}),
 b_{2g}(3d_{xy}), e_{g}(3d_{xz},3d_{yz})$
 orbitals hybridize, due to strong Cu $3d$-O $2p$ covalency, with even O
$2p$-orbitals
 of  the same symmetry, thus forming appropriate bonding
$\gamma ^{b}$ and antibonding $\gamma ^{a}$ states. Among the odd
orbitals only $e_{u}(\sigma)$ and $e_{u}(\pi)$ hybridize due to
nearest neighbor $pp$ overlap and transfer thus forming appropriate
bonding $e_{u}^{b}$ and antibonding $e_{u}^{a}$ purely oxygen
states. The purely oxygen $a_{2g},a_{2u},b_{2u}$ orbitals are
nonbonding.
\begin{figure}[t]
\includegraphics[width=8.5cm,angle=0]{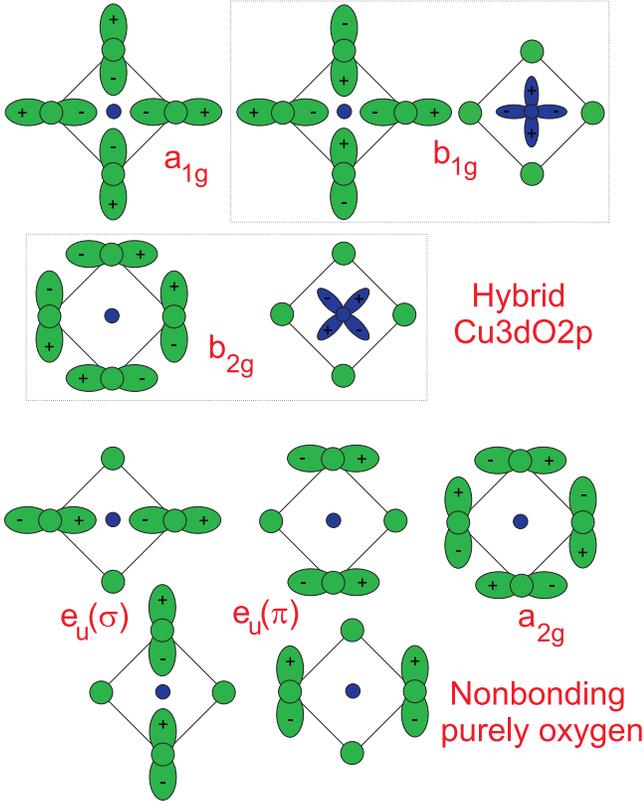}
\caption{Electron (hole) density distribution for planar copper
and oxygen molecular orbitals of one CuO$_4$ plaquette.}
\label{fig1}
\end{figure}
All "planar" O $2p$ orbitals (see Fig.\ \ref{fig1}) in accordance
with the orientation of lobes could be classified as $\sigma$
($a_{1g},b_{1g},e_{u}(\sigma)$) or $\pi$
($a_{2g},b_{2g},e_{u}(\pi)$) orbitals, respectively.

Bonding and antibonding molecular orbitals in hole representation
can be presented as
$$
|\gamma ^{b}\rangle =\cos \alpha _{\gamma} |\gamma (3d)\rangle +
\sin \alpha _{\gamma}|\gamma (2p)\rangle ,
$$
$$
|\gamma ^{a}\rangle =\sin \alpha _{\gamma} |\gamma (3d)\rangle -
\cos \alpha _{\gamma}|\gamma (2p)\rangle .
$$
For example,
$$
|b_{1g}^{b}\rangle =\cos \alpha _{b_{1g}} |b_{1g}(3d)\rangle + \sin
\alpha _{b_{1g}}|b_{1g} (2p)\rangle ,
$$
\begin{equation}
| b_{1g}^{a}\rangle =\sin \alpha _{b_{1g}} |b_{1g}(3d)\rangle - \cos
\alpha _{b_{1g}}|b_{1g}(2p)\rangle ,
\end{equation}
where $b_{1g}(3d)=3d_{x^2-y^2}$;
$$
|e_{u}^{b}\rangle =\cos \alpha _{e}\,|e_{u}(\pi)\rangle + \sin
\alpha _{e}\,|e_{u}(\sigma)\rangle ,
$$
\begin{equation}
|e_{u}^{a}\rangle =\sin \alpha _{e}\,|e_{u}(\pi)\rangle - \cos
\alpha _{e}\,|e_{u}(\sigma)\rangle
\end{equation}
equally for both types ($x,y$) of such orbitals.

To explain the electronic structure of the embedded molecular
cluster in a broad energy range we need the complete information
about the bare parameters of the effective Hamiltonian.
The numerical values of energy parameters for holes in insulating
cuprates like La$_2$CuO$_4$ or Sr$_2$CuO$_2$Cl$_2$ which were used
by different authors \cite{Stechel,McM,Czyzyk,Mattheiss,Hayn} can
be summarized  as follows (in eV):
$$
\epsilon _{p}-\epsilon _{d}\approx 2
$$
$$
\epsilon _{p\sigma}-\epsilon _{p\pi}\approx 1\div 3 ;\quad \epsilon
_{p\pi}<\epsilon _{pz}<\epsilon _{p\sigma};
$$
$$
U_{d}\approx 6\div 10\, eV; \quad U_{p}\approx 4\div 6; \quad
V_{pd}\approx 0.5\div 1.5;
$$
$$
t_{pd}\approx 1.0\div 1.5; \quad |t_{pp\sigma}|\approx 0.8 ; \quad
t_{pp\pi}\approx \frac{1}{2}|t_{pp\sigma}|\approx 0.4 \,,
$$
where
$$
\epsilon _{d}=\frac{1}{5}\sum _{i}\epsilon _{d,i}=0;\quad \epsilon
_{p}=\frac{1}{3}(\epsilon _{p\sigma}+\epsilon _{p\pi}+\epsilon
_{pz})
$$
are the centers of "gravity" for the Cu $3d$ and O $2p$ manifolds,
respectively. As will be discussed more in detail below (see Sec.\
VI.E.1) such a parameter choice is confirmed by the corresponding
ARPES data \cite{Pothuizen,Duerr} giving important information
concerning the nonbonding oxygen states.
\begin{figure}[t]
\includegraphics[width=8.5cm,angle=0]{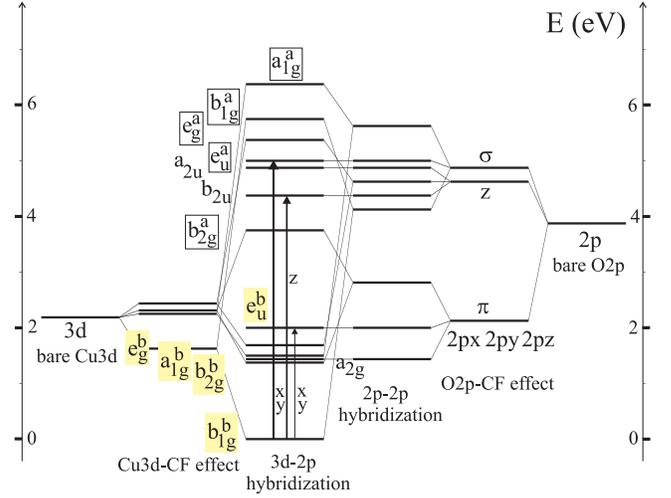}
\caption{Model single-hole energy spectra for a CuO$_4$ plaquette
with parameters relevant for Sr$_2$CuO$_2$Cl$_2$ and a number of
other insulating cuprates.} \label{fig2}
\end{figure}
Fig. \ref{fig2}  presents a possible single-hole energy spectrum
for a CuO$_4$ plaquette embedded into an insulating cuprate like
Sr$_2$CuO$_2$Cl$_2$ calculated with the parameters above. For
illustration we show also a  step-by-step formation of the cluster
energy levels from the bare Cu $3d$ and O $2p$ levels with the
successive inclusion of crystalline field (CF) effects, O $2p$-O
$2p$, and Cu $3d$-O $2p$ covalency.

\section{One-center Frenkel-like excitons}

\begin{figure}[t]
\includegraphics[width=8.5cm,angle=0]{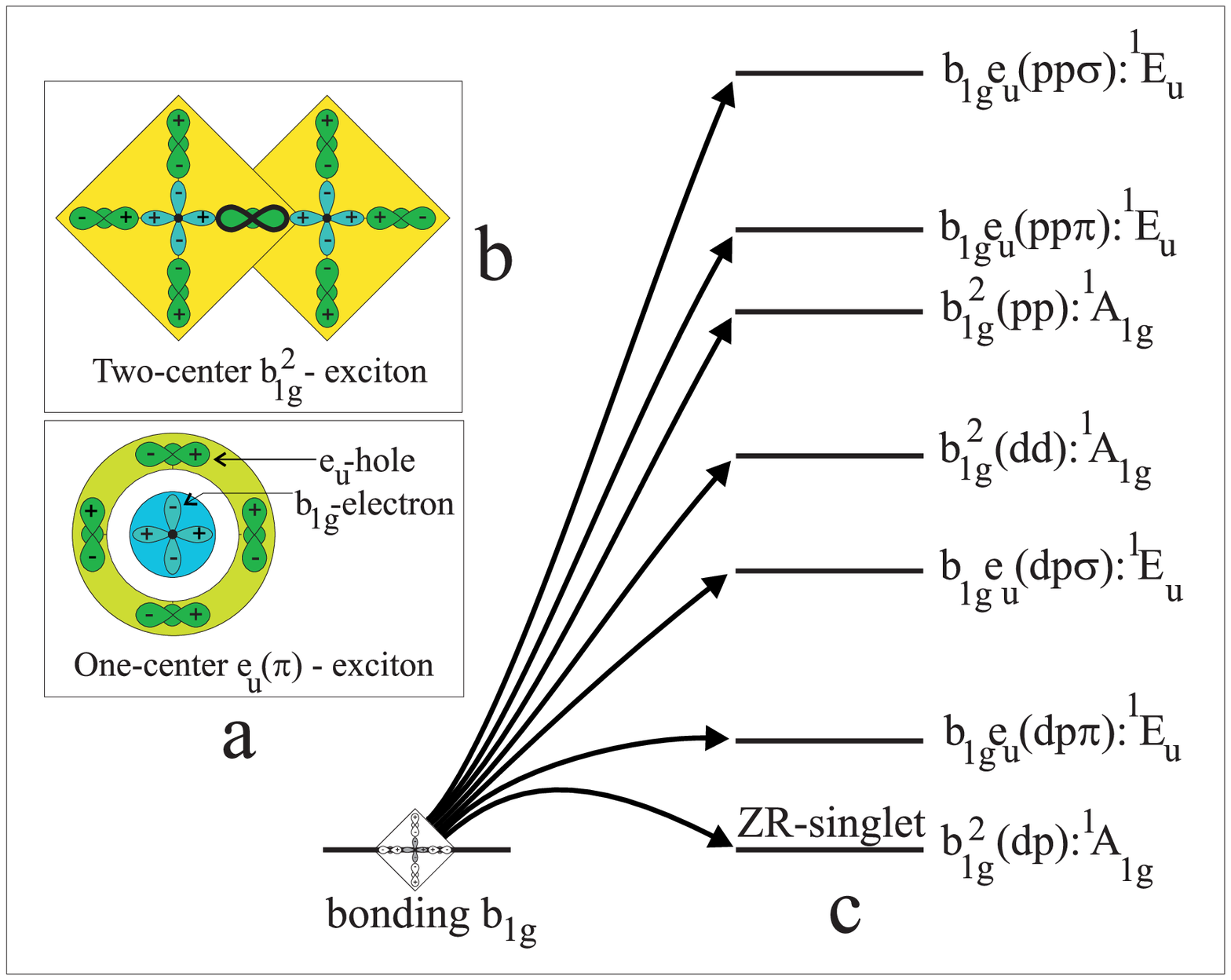}
\caption{(a) Simplified illustration of the one-center exciton and
(b) the two-center exciton of $b_{1g}^2;pd$, i.e.\  ZR-singlet
type. (c) Classification of the important two-center excitons
together with an energy scheme.} \label{fig3}
\end{figure}

Small charge transfer excitons, or excited states arising from the
configuration in which an electron is transferred from a negative
ion to a nearest neighbor positive ion have been considered many
years ago to be the origin of excitonic-like peaks which show up
as a low-energy structure of the fundamental absorption band, for
instance in alkali halide crystals. \cite{Overhauser} Namely this
idea was exploited by Zhang and Ng in their simple model theory of
small CT excitons in insulating cuprates. \cite{Wang,Ng} However,
in terms of the CuO$_4$ cluster model the local two-atomic Cu
$3d$- O $2p$ charge transfer generates a number of both
intra-cluster and inter-cluster transitions (see Fig.\
\ref{fig3}).
The corresponding electron-hole excitations form small one- and
two-center excitons, respectively. The main criteria are the
binding energy and the average electron-hole separation. Namely
the latter provides the localization and stability of excitations
due to their small overlap for nearest neighbors (small tunneling
amplitude), and finally leads to a narrow exciton band. Below we
address the most probable candidate states for small one- and
two-center CT excitons.  It should be noted that the main concept
which we apply  has in many aspects a close similarity with those
used for CT excitons of linear-chain organic $\pi$-conjugated
polymers. \cite{Guo,Boman} Some symmetry aspects of the CT
excitons in cuprates were addressed by Cherepanov {\it et al}.
\cite{Cherepanov}

\subsection{Classification}

Among the numerous one-center electron-hole CT excitations the first
candidates for dipole-active excitons are CT transitions $b_{1g}^{b}
\to e_{u}^{a,b}$ from the ground state $b_{1g}^{b}$ to the purely
oxygen doublet O $2p_\pi $-O $2p_\sigma$ hybrid states
$e_{u}^{a,b}$, which are allowed in "in-plane" polarization ${\bf
E}\perp C_{4}$, and the $b_{1g}^{b}\to b_{2u}$ transition to purely
oxygen  O $2p_z$-like state, which is allowed in "out-of-plane"
polarization ${\bf E}\parallel C_{4}$. All these one-center excitons
may be rather simply represented as a hole rotating on the four
nearest oxygens around an electron predominantly localized in the Cu
$3d_{x2-y2}$ state with a minimal electron-hole separation
$R_{eh}\approx R_{CuO}\approx$ 2\AA \ (see Fig.\ \ref{fig3}).

The two excitons $b_{1g}^{b} \to e_{u}^{b}$ and $b_{1g}^{b} \to
e_{u}^{a}$  differ by the oxygen hole density distribution: for
the former this has a predominantly O $2p_\pi$ character, while
for the latter it has a  O $2p_\sigma$ one. It should be noted
that the excitonic doublet state may be considered as two current
states $e_{u}(\pm 1)$, or two currentless states $e_{u}(x,y)$ with
quenched orbital motion. The former are excited by circular
polarized light, while the latter by linear polarized one. The
double degeneracy for the $b_{1g}^{b} \rightarrow e_{u}^{b}$
excitons results in anomalously strong electron-lattice coupling
with all features typical for the Jahn-Teller
effect,\cite{bersuker} in particular, high probability to form a
self-trapped state. In any case, these excitons should be
considered as first candidates for resonant phonon Raman
scattering activity. One should specifically emphasize that along
with resonance excitation of the allowed Raman modes the
Jahn-Teller excitons could generate forbidden phonon modes.

 Among dipole inactive excitons one should note $b_{1g}^{b} \to b_{2g}$ and,
especially,
 $b_{1g}^{b} \to a_{2g}$  with purely oxygen O $2p_\pi$  holes. The latter
incorporates a purely oxygen $a_{2g}$ hole, which has in frames of
our model the minimal energy among all the purely oxygen states. In
accordance with the model energy spectrum (see Ref.\
\onlinecite{PRB}) its energy should be of the order of $1.5\div 1.8$
eV, or in other words appears to be lower than the optical gap. This
circumstance draws specific attention to this exciton, despite the
$b_{1g}^{b} \to a_{2g}$ transition is dipole forbidden. Due to the
purely oxygen nature of the hole, the  $b_{1g}^{b} \to a_{2g}$
exciton  resembles the dipole active  $b_{1g}^{b} \to e_{u}$
exciton, however, contrary to the latter this is a purely  O
$2p_\pi$ hole.

There are several examples of a qualitative assignement of low-lying
optical spectral features to the non-bonding oxygen orbitals,
appearing in CuO near 1.7 eV (see Ref.\
\onlinecite{Moskvin1994,Sukhorukov} and in Sr$_2$CuO$_2$Cl$_2$ near
2.5 eV (see Ref.\ \onlinecite{Choi}). Also, it was shown that the
dipole-allowed one-center electron-hole excitation $b_{1g}^{b} \to
e_{u}(\pi)$ is visible in the low lying part of the EELS spectrum of
Sr$_2$CuO$_2$Cl$_2$. \cite{PRB} However, to the best of our
knowledge, there is at present no unambiguous identification of this
one-center excitation (and also of $b_{1g} \to b_{2u}$) as
well-separated entities in 2D insulating cuprates. That is different
for 1D cuprates where these one-center excitations built up of
non-bonding oxygen-orbitals could be separated in EELS by choosing a
momentum transfer perpendicular to the chain direction. \cite{PRL}
The conventional one- and two-band Hubbard-like models consider only
the Cu $3d_{x^2-y^2}$ and O $2p_{\sigma}$ orbitals, and do not take
into account the nonbonding O $2p_{\pi}$ and O $2p_z$ orbitals.
Partly, that may be justified by the specific properties of the
$b_{1g}$ ($d_{x^2-y^2}$) ground state resulting in a predominant
contribution of O $2p_{\sigma}$ orbitals both to the Cu $3d$ - O
$2p$ bonding and to the different charge transfer transitions.
Optical absorption and EELS activity associated with the nonbonding
O $2p_{\pi}$ and O $2p_z$  orbitals appear to be rather weak if any.

\subsection{Transition matrix elements in optics and EELS}

\subsubsection{ Intensities of the electric dipole CT  transitions for
one-center small  excitons}

Electric dipole CT transitions for the hole localized in the CuO$_4$
cluster are allowed from the even ground state $b_{1g}^b$ to the
bonding and antibonding purely oxygen odd $e_{u}^{a,b}$ states for
the ${\bf E}\perp C_4$ polarization, and to purely oxygen odd
$b_{2u}$ state for the ${\bf E}\parallel C_4$ one (see Fig.\
\ref{fig1}). Dipole transition matrix elements for the $b_{1g}^b$
ground state depend essentially  on the fact whether the final state
is of O $2p_\pi$ or O $2p_\sigma$ character. For the former case we
have only a "nonlocal overlap contribution" with two-center
integrals
$$
\langle b_{1g}^b|qx|e_{ux}(\pi)\rangle =-\langle
b_{1g}^b|qy|e_{uy}(\pi)\rangle
$$
$$
= \cos\alpha _{b_{1g}}\langle d_{x^2 -y^2}|qx|e_{ux}(\pi)\rangle
$$
$$
 =q\sqrt{2} \cos\alpha _{b_{1g}}
 \int \phi _{d_{x^2 -y^2}}({\bf r})^* x\phi _{px}({\bf r}-{\bf R})d{\bf r};
$$
\begin{equation}
 \langle b_{1g}^b|qz|b_{2u}\rangle = q\sqrt{2} \cos\alpha _{b_{1g}}
 \int \phi _{d_{x^2 -y^2}}({\bf r})^* z\phi _{pz}({\bf r}-{\bf R})d{\bf r},
 \label{dipole}
\end{equation}
 where ${\bf R}\parallel {\bf y}$, and
$$
\int \phi _{d_{x^2 -y^2}}({\bf r})^* x\phi _{px}({\bf r}-{\bf
R})d{\bf r}
$$
$$
\qquad \qquad = -\int \phi _{d_{x^2 -y^2}}({\bf r})^* z\phi
_{pz}({\bf r}-{\bf R})d{\bf r},
 $$
 that results in equal intensities for $b_{1g}^{b}\rightarrow e_{u}(\pi)$ and
 $b_{1g}^{b}\rightarrow b_{2u}$ transitions. Naturally, that in a more common
case we have
 orthogonalized Wannier functions instead of simple atomic ones, and the
calculation of
 the nonlocal overlap matrix elements is not so straightforward.
 For the $b_{1g}^{b}\rightarrow e_{u}(\sigma)$ dipole transitions to purely
 O $2p_\sigma$
 state we have both nonlocal and local ( or "covalent") overlap  contributions
  due to nonzero O $2p_\sigma$
 density in the initial $b_{1g}^{b}$ state
 $$
\langle b_{1g}^b|qx|e_{ux}(\sigma)\rangle =-\langle
b_{1g}^b|qy|e_{uy}(\sigma)\rangle
 $$
 $$
 = q \sqrt{2}\cos\alpha _{b_{1g}}
 \int \phi _{d_{x^2 -y^2}}({\bf r})^* y\phi _{py}({\bf r}-{\bf R})d{\bf r}
$$
$$
 +
\frac{q R_{CuO}}{\sqrt{2}} \sin\alpha _{b_{1g}} \; ,
$$
 where ${\bf R}\parallel {\bf y}$.
  It should be noted that namely the local term is the only contribution
usually addressed
 for transition matrix elements.

 The bonding and antibonding  $e_{u}^{a,b}$ states incorporate both O
$2p_\sigma$
 and O $2p_\pi$ orbitals. Therefore,  the respective dipole transition matrix
elements
 acquire a rather complex structure. However, for the "covalent", and probably,
 the leading contribution, we obtain a simple relation
 \begin{equation}
\frac{I(b_{1g}^{b} \to e_{u}^{b})}{I(b_{1g}^{b} \to
e_{u}^{a})}=|\tan \alpha _{e_u}|^2 \approx
\left|\frac{t_{pp\sigma}+t_{pp\pi}}{\epsilon _{pe_{u}(\sigma )}-
\epsilon _{pe_{u}(\pi )}} \right|^2 \; .
\end{equation}
 In other words, the relative intensity of the two main electric-dipole transitions
 is mainly determined by the magnitude of $pp$-hybridization and does not
exceed 0.1 for the typical  values of the parameters.

 In some cases, it is of practical interest to provide information concerning
the
 allowed dipole transitions between excited states. In our case,
 these are, first of all, numerous transitions from or to $e_u$ states. For instance, the
 $a_{2g}\rightarrow e_{u}^b$ dipole allowed transition has a very large
 intensity due to a large local contribution to the matrix element:
 $$
 \langle a_{2g}|qx|e_{ux}(\pi)\rangle = \frac{q R_{CuO}}{\sqrt{2}} \; .
 $$
  It should be noted that the
 rather simple model of a single CuO$_4$ cluster allows already to
 predict both the energies and the relative intensities for such transitions.

\subsubsection{ Intensities of the EELS transitions for one-center small
excitons}

The intensity of an exciton is determined by the imaginary part of
$\epsilon({\bf k},\omega)$ around a pole, and is given by
\begin{equation}
I\propto k^{-2}|\langle\Psi _{exc}|e^{i{\bf k}{\bf r}}|\Psi
_{GS}\rangle|^2 ,
\end{equation}
where $\Psi _{exc}$ is the exciton wave function. \cite{Ng} Within
the long-wavelength approximation ($k \rightarrow 0$) the EELS
selection rules are the same as in optics if we address the ${\bf
k}$ direction to be that of the electric field. For the one-center
small $\gamma$-excitons it is rather easy to obtain the angular
${\bf k}$-dependence of the transition matrix elements in the
small ${\bf k}$-approximation:
\begin{equation}
 \langle\Psi _{exc}|e^{i{\bf k}{\bf r}}|\Psi _{GS}\rangle \propto
  \langle\gamma|e^{i{\bf k}{\bf r}}|b_{1g}^{b}\rangle \propto k^L
Y_{L\gamma}({\bf k}),
\end{equation}
where $Y_{L\gamma}({\bf k})$ is a linear superposition of spherical
harmonics $Y_{LM}({\bf k})$, which forms a basis of the irreducible
representation $\gamma$ of the $D_{4h}$ point group. We make use of
the familiar expansion
\begin{equation}
e^{i{\bf
k}{\bf{r}}}=4\pi\sum_{L=0}^{\infty}\sum_{M=-L}^{L}i^{L}j_{L}(kr)Y_{LM}({\bf
k}) Y^{*}_{LM}({\bf{r}}) \label{exp},
\end{equation}
 restricting ourself to the first nonzero contribution with minimal $L$.
Thus, for different $\gamma$ one obtains:
 $$
 \langle e_{u}|e^{i{\bf k}{\bf r}}|b_{1g}^{b}\rangle \propto k Y_{1e_u}({\bf
 k}) \propto k_{x,y};(\mbox{dip.\ approx.})
 $$
$$
 \langle a_{1g}|e^{i{\bf k}{\bf r}}|b_{1g}^{b}\rangle \propto k^2
Y_{2b_{1g}}({\bf k})
 \propto (k_{x}^{2}-k_{y}^{2}); (\mbox{quadr.\ approx.})
 $$
 $$
 \langle a_{2g}|e^{i{\bf k}{\bf r}}|b_{1g}^{b}\rangle \propto k^2
Y_{2b_{2g}}({\bf k})
 \propto k_{x}k_{y}; (\mbox{quadr.\ approx.})
 $$
 $$
 \langle b_{1g}|e^{i{\bf k}{\bf r}}|b_{1g}^{b}\rangle \propto k^4
Y_{4a_{1g}}({\bf k})
 \propto \cos 4\phi; (\mbox{oct.\ approx.})
 $$
 $$
 \langle b_{2g}|e^{i{\bf k}{\bf r}}|b_{1g}^{b}\rangle \propto k^4
Y_{4a_{2g}}({\bf k})
 \propto \sin 4\phi; (\mbox{oct.\ approx.})\,.
 $$
Here, $\phi$ is the azimuthal angle of orientation for the ${\bf
k}$ vector. In general,  among dipole inactive excitons for
corner-sharing CuO$_4$ plaquette systems the $a_{1g}$ and $a_{2g}$
ones are "visible" only along $[100]$ and $[110]$ directions,
respectively; the $b_{1g}^{a}$ exciton is visible both along
$[100]$ and $[110]$ directions, while the $b_{2g}$ one does not
manifest itself both along $[100]$ and $[110]$ directions. Taking
into account only the conventional local overlap ("covalent")
contribution to the transition matrix element one may obtain
model expressions for $\langle \gamma|e^{i{\bf k}{\bf
r}}|b_{1g}^{b}\rangle $ suitable throughout the BZ. Thus, one
obtains for the O $2p$ local overlap contribution:
$$
\langle a_{2g}|e^{i{\bf k}{\bf r}}|b_{1g}^{b}\rangle = \langle
b_{2g}|e^{i{\bf k}{\bf r}}|b_{1g}^{b}\rangle=0;
$$
$$
\langle a_{1g}^{a}|e^{i{\bf k}{\bf r}}|b_{1g}^{b}\rangle
=-\frac{1}{2} \sin \alpha _{b_{1g}}\cos \alpha _{a_{1g}}\left[
\cos \frac{k_{x}a}{2}- \cos\frac{k_{y}a}{2}\right];
$$
\begin{equation}
\langle b_{1g}^{a}|e^{i{\bf k}{\bf r}}|b_{1g}^{b}\rangle
=-\frac{1}{2} \sin \alpha _{b_{1g}}\cos \alpha _{b_{1g}}\left[\cos
\frac{k_{x}a}{2}+ \cos \frac{k_{y}a}{2}\right];
\end{equation}
$$
\langle e_{u}^{b}x|e^{i{\bf k}{\bf r}}|b_{1g}^{b}\rangle
=\frac{i}{\sqrt{2}} \sin \alpha _{b_{1g}}\sin \alpha _{e}\sin
\frac{k_{x}a}{2},
$$
$$
\langle e_{u}^{b}y|e^{i{\bf k}{\bf r}}|b_{1g}^{b}\rangle
=-\frac{i}{\sqrt{2}} \sin \alpha _{b_{1g}}\sin \alpha
_{e}\sin\frac{k_{y}a}{2};
$$
$$
\langle e_{u}^{a}x|e^{i{\bf k}{\bf r}}|b_{1g}^{b}\rangle
=-\frac{i}{\sqrt{2}} \sin \alpha _{b_{1g}}\sin \alpha
_{e}\cos\frac{k_{x}a}{2},
$$
$$
\langle e_{u}^{a}y|e^{i{\bf k}{\bf r}}|b_{1g}^{b}\rangle
=-\frac{i}{\sqrt{2}} \sin \alpha _{b_{1g}}\cos \alpha
_{e}\sin\frac{k_{y}a}{2}.
$$
For the dipole-allowed $b_{1g}\rightarrow e_u$ transitions  the EELS
intensity has to decrease in going from the $\Gamma$-point to the BZ
boundary with
$$
\frac{I(\pi /a,\pi /a,0)}{I(0,0,0)}\,=\, \frac{I(\pi
/a,0,0)}{I(0,0,0)}=\frac{4}{\pi ^2}\approx 0.4 .
$$
One should note that with the inclusion of the copper contribution
the final expression for the matrix element $\langle
b_{1g}^{a}|e^{i{\bf k}{\bf r}}|b_{1g}^{b}\rangle $ should be
modified
$$
\langle b_{1g}^{a}|e^{i{\bf k}{\bf r}}|b_{1g}^{b}\rangle
$$
\begin{equation}
\approx -\frac{1}{2} \sin \alpha _{b_{1g}}\cos \alpha
_{b_{1g}}\left[\cos\frac{k_{x}a}{2}+
\cos\frac{k_{y}a}{2}-2\right].
\end{equation}
Interestingly, that in the frames of the local model for
transition matrix elements the $a_{2g}$ and $b_{2g}$ one-center
excitons appear to be invisible throughout the BZ body.

\section{Two-center excitons}

\subsection{Electronic structure and classification}

Inter-center charge transfer transitions between two CuO$_4$
plaquettes centered at neighboring sites A and B define two-center
excitons in the molecular cluster Cu$_2$0$_7$. These two-center
excitons may be considered as quanta of the disproportionation
reaction
\begin{equation}
\mbox{CuO}_{4}^{6-}+\mbox{CuO}_{4}^{6-}\rightarrow
\mbox{CuO}_{4}^{7-}+ \mbox{CuO}_{4}^{5-}
\end{equation}
with the creation of electron  CuO$_{4}^{7-}$ and hole
CuO$_{4}^{5-}$ centers. The former corresponds to completely
filled Cu $3d$ and O $2p$ shells, or the vacuum state for holes,
while the latter may be found in different two-hole states. These
two-hole states can be classified by $\gamma_1 \gamma_2=\Gamma$
according to the symmetries $\gamma_i$ of each of the holes. For
instance, if we restrict ourselves to charge transfer processes
governed by the strongest $\sigma$ bond, one has to distinguish
three different channels $b_{1g}^2$, $b_{1g}a_{1g}$ and
$b_{1g}e_{u}$. Only two of them ($b_{1g}^2$ and $b_{1g}e_{u}$)
have non zero optical matrix elements, but all are important for
EELS. Each symmetry different channel contains several wave
functions with different Cu and O character (see Section IV.A.1).

To construct ground state and charge transfer excited states we
use an approach similar to the well known Heitler-London scheme.
In this approach the two hole states for the Cu$_2$O$_7$ molecular
cluster are formed from different one-center states. We deal with
charge states of the CuO$_4$ plaquette with zero, one, and two
holes, respectively, denoted as $\Phi _{A,B}^{(0,1,2)}$. Then, the
ground state $\Psi_{GS}$ of the cluster Cu$_2$0$_7$ is
predominantly given by the symmetrized product of
$\Phi_{A}^{(1)}(b_{1g}^{b}) \Phi_{B}^{(1)}(b_{1g}^{b})$ coupled to
higher states of the same symmetry. Accordingly, the excited
(exciton) states are built by product states
\begin{equation}
\phi_{ES}^{eh}(\Gamma)=\Phi_{A}^{(0)} \Phi_{B}^{(2)}(\Gamma) \; ,
\label{eh1}
\end{equation}
 with subsequent (anti)symmetrization
(see Section IV.A.2).

\subsubsection{Two-hole configurations for the CuO$_{4}^{5-}$
center}

To classify all two-hole configurations of one CuO$_{4}^{5-}$
center we use their symmetry $\Gamma=\gamma_1 \gamma_2$ and take
into account the Coulomb interaction. The latter is of particular
interest both for energetics and the electron structure of
two-center excitons. According to symmetry we have to distinguish
even $\Gamma$ configurations (like $b_{1g}^{2}$ or $b_{1g}
a_{1g}$) and odd $\Gamma$ configurations (as $b_{1g} e_{u}$).
Alternatively to the notation $\gamma_1 \gamma_2$ we will also use
the representations of the $D_{4h}$ symmetry of the CuO$_4$
cluster (see Ref.\ \onlinecite{Eskes}) with the following
correspondence: $b_{1g}^{2}={}^{1}A_{1g}$ and $b_{1g}
e_{u}={}^{1}E_u$.  Each channel of two-center excitons,
characterized by a $\gamma_1 \gamma_2$ configuration, consists of
several wave functions differing by their partial Cu 3$d$ and O
2$p$ density distribution. These excitons will be denoted by
$\gamma_1 \gamma_2;dd$, $\gamma_1 \gamma_2;pd$, and
$\gamma_1\gamma_2;pp$, respectively. In the sense of quantum
chemistry the mixing between several wave functions of one channel
can be understood as a configuration interaction effect.

Let us present the simple example of the calculation of the
two-hole spectrum in the Zhang-Rice (ZR)-singlet sector \cite{ZR}
(i.e.\ in the $b_{1g}^2$-channel). As usually, we assume that the
ZR-singlet represents the lowest spin-singlet state formed by the
interaction of three "purely ionic" two-hole configurations
$|d^{2}\rangle$, $|pd\rangle$, and $|p^{2}\rangle$. Here,
$|d\rangle=|d_{x^2 -y^2}\rangle$ and
$|p\rangle=|p_{b_{1g}}\rangle$ are the non-hybridized Cu 3$d_{x^2
-y^2}$ and O 2$p_{\sigma}$ orbitals, respectively, with bare
energies $\epsilon_{d}$ and $\epsilon_{p}$. Then, the matrix of
the full effective Hamiltonian within the bare basis set has  a
rather simple form
\begin{equation}
{\hat H} =\pmatrix{2\epsilon _{d}+U_{d} & t & 0 \cr t & \epsilon
_{d}+\epsilon _{p}+V_{pd} & t \cr 0 & t & 2\epsilon _{p}
+U^{*}_{p}\cr}, \label{U}
\end{equation}
where the effective Coulomb parameter for purely oxygen
configuration incorporates both the intra-atomic parameter $U_p$
and the oxygen-oxygen coupling to the first and second nearest
neighbors, respectively
$$
U^{*}_{p}=U_{p}+ \frac{1}{4}V_{pp}^{(1)}+\frac{1}{8}V_{pp}^{(2)},
$$
and the following condition holds:
$$ U_d \, > \,U_p \, >\, V_{pd}
$$
For reasonable values of parameters (in eV): $U_{d}= 8.5$,
$U_{p}=4.0$, $V_{pd}=1.2$, $\epsilon _{d}=0$, $\epsilon _{p}=3.0$,
$t=t_{pd}=1.3$ (see Ref.\  \onlinecite{Hybertsen}) we obtain for
the ZR-singlet energy $E_{ZR}=3.6$, and its wave function
\begin{equation}
|\Phi_1^{(2)}\rangle=|b_{1g}^{2};pd\rangle = -0.25|d^{2}\rangle
+0.95|pd\rangle  -0.19|p^{2}\rangle \; . \label{zr1}
\end{equation}
It reflects the well-known result that the ZR-singlet represents a
two-hole configuration with one predominantly Cu $3d$ and one
predominantly O $2p$ hole. The two excited states with energy
$E_{ZR}+5.2$ and $E_{ZR}+6.7$ eV are described by the wave
functions
\begin{equation}
|\Phi_2^{(2)}\rangle=|b_{1g}^{2};dd\rangle = 0.95|d^{2}\rangle
+0.21|pd\rangle -0.22|p^{2}\rangle , \label{zr2}
\end{equation}
\begin{equation}
|\Phi_3^{(2)}\rangle=|b_{1g}^{2};pp\rangle = -0.17|d^{2}\rangle
-0.24|pd\rangle -0.96|p^{2}\rangle , \label{zr3}
\end{equation}
respectively. Given the ZR-singlet energy one may calculate the
transfer energy from $b_{1g}^{b}$ to the neighboring $b_{1g}^2;pd$
state (the ZR-singlet state), or in short the $b_{1g}^{b} \to
b_{1g}$ transfer energy:
$$
\Delta_{CT}=E_{ZR}-2E_{b_{1g}}=(3.6 +0.5) \mbox{eV} = 4.1
\mbox{eV},
$$
where the stabilization energy for the bonding $b_{1g}^b$ state is
simply calculated from matrix (\ref{U}) at
$U_{d}=U^{*}_{p}=V_{pd}=0$. One should note that  the minimal
$b_{1g}^{b} \to b_{1g}$ transfer energy relatively weakly depends
on the value of the $pd$ transfer integral; we obtain the same
value $4.2$ eV at any value of $t$ between $1.0\div 1.5$ eV. This
energy depends mainly on the values of
$(\epsilon_{p}-\epsilon_{d})$ and $V_{pd}$.

Similar charge transfer energies are obtained in various model
approaches. It should be emphasized that this quantity plays a
particular role as the minimal charge transfer energy which
specifies the charge transfer gap.  In the general case it is
defined
$$
\Delta _{CT}= E_{N+1}+E_{N-1}-2E_{N},
$$
as the energy required to remove a hole from one region of the
crystal and add it to another region beyond the range of excitonic
correlations. The exact diagonalization studies for a series of
clusters with different size \cite{Hybertsen,Hybertsen1} show that
$\Delta _{CT}$ strongly diminishes with cluster size from $\approx
4$eV for small clusters to $\approx 2.5$eV as extrapolated value
for large clusters.

\subsubsection{Ground state and charge transfer excited
states of the Cu$_2$O$_7$ cluster}

As already noted, the states $\Phi _{A,B}^{(0,1,2)}$ of the
CuO$_4$ plaquette centered at A or B will be used to construct the
Cu$_2$O$_7$ wave functions in a Heitler-London like scheme.
Moreover, for two-center molecular clusters like Cu$_2$O$_7$ with
$D_{2h}$ point symmetry  one has two types of excitations:
$\phi_{ES}^{eh}(\Gamma)=\Phi_{A}^{0}\Phi_{B}^{2}(\Gamma)$, and
$\phi_{ES}^{he}(\Gamma)=\Phi_{A}^{2}(\Gamma)\Phi_{B}^{0}$, which
differ only by the permutation of electron and hole. These
functions will interact due to the resonance reaction
\begin{equation}
\mbox{CuO}_{4}^{7-}+\mbox{CuO}_{4}^{5-} \rightarrow
\mbox{CuO}_{4}^{5-} +\mbox{CuO}_{4}^{7-} \; ,
\label{r2}
\end{equation}
giving rise to the splitting of the two bare excitations discussed
in more detail below.

The spin-singlet even ground state for the Cu$_2$O$_7$ system can
be written as follows
 \begin{equation}
\Psi_{GS}= \cos\alpha \,\phi^{g}_{b_{1g}^{b}b_{1g}^{b}}+
\sin\alpha \sum _{\Gamma} a_{\Gamma}\phi^{g}_{ES}(\Gamma),
\label{GS2}
\end{equation}
where $\phi^{g}_{b_{1g}^{b}b_{1g}^{b}}$ is the symmetrized product
of the one-center one-hole states $\Phi _{A}^{(1)}(b_{1g}^{b})
\Phi _{B}^{(1)}(b_{1g}^{b})$. It is coupled to the even states
\begin{equation}
\label{gd}
 \phi^{g}_{ES}(\Gamma)=
\frac{1}{\sqrt{2}}(\phi_{ES}^{eh}(\Gamma)\pm
\phi_{ES}^{he}(\Gamma))
\end{equation}
corresponding to symmetrized (antisymmetrized) bare states
$\phi_{ES}^{eh}(\Gamma)$ for even (odd) $\Gamma$. Correspondingly,
the odd and even excited CT states are given by
$$
\Psi^{u}_{ES}=\sum _{\Gamma}b_{\Gamma}\phi^{u}_{ES}(\Gamma);
$$
\begin{equation}
\Psi^{g}_{ES}= \cos\alpha \sum
_{\Gamma}c_{\Gamma}\phi^{g}_{ES}(\Gamma) - c_{ES}\sin\alpha
\phi^{g}_{b_{1g}^{b}b_{1g}^{b}}, \label{odd1}
\end{equation}
where the index $\Gamma$ labels different final configurations of
the  two-hole CuO$_4$ cluster denoted by $\Gamma=\gamma_{1}
\gamma_{2};dd$, $\gamma_{1} \gamma_{2}; pd$, and
 $\gamma_{1} \gamma_{2};pp$, respectively. In
close analogy  to (\ref{gd}) we can write
\begin{equation}
\label{ud}
 \phi^{u}_{ES}(\Gamma)=
\frac{1}{\sqrt{2}}(\phi_{ES}^{eh}(\Gamma)\mp
\phi_{ES}^{he}(\Gamma));
\end{equation}
for even (odd) $\Gamma$.

\subsubsection{The $S$- and $P$-like two-center excitons}

The energies of the superposition states (\ref{gd}) and (\ref{ud})
are given by $E_0 \pm |T_{AB}|$, where $E_0$ is the energy of the
bare state $\phi_{ES}^{eh}(\Gamma)$ and $T_{AB}$ is an effective
resonance two-particle transfer integral of the resonance reaction
(\ref{r2}). The even (odd) states $\phi_{ES}^{g}$
($\phi_{ES}^{u}$) may also be denoted as $A_g$ ($B_{1u}$) states
corresponding to the $D_{2h}$ point symmetry of the Cu$_2$O$_7$
cluster. Alternatively, they correspond to $S$- (for $A_g$) or
$P$-like (for $B_{1u}$) two-center excitons. Namely these $S$- or
$P$-like excitons will draw our main attention below. For brevity,
we use the simple $g,u$ labels instead of the $A_g$, $B_{1u}$
notations.

The different coefficients in (\ref{GS2}) and (\ref{odd1}) are
coupled by normalization and orthogonality conditions like
$$
 \sum _{\Gamma}|a_{\Gamma}|^2 = \sum _{\Gamma}|b_{\Gamma}|^2 =1; \,
 c_{ES}=\sum _{\Gamma}a^{*}_{\Gamma}c_{\Gamma}.
$$
Its magnitude is actually determined by appropriate hole transfer
integrals. It should be noted, however, that our approach is only an
approximative one. So, due to the small overlap between the two
neighboring CuO$_4$ plaquettes, the one-hole $\gamma$ label is
likely to be a "bad quantum number". Nevertheless, it can be used in
the frames of a semi-quantitative approach. In the single channel
approximation where the exciton is dominated by the transfer from
$b_{1g}^{b}$ to one neighboring $b_{1g} \gamma$ state, we may write
\begin{equation}
\sin\alpha = \frac{t_{b_{1g} \to \gamma}}{\Delta_{b_{1g} \to
\gamma}} ,
\end{equation}
where $t_{b_{1g} \to \gamma}$ and $\Delta_{b_{1g} \to \gamma}$ are
the CT transfer integral and CT energy, respectively.

Let us note that in our approach the $S$- and $P$-excitons are
centered at the central oxygen ion of the Cu$_2$O$_7$ cluster.
That is in contrast to the simple ZN-model \cite{Ng} which implies
the hole location on one of the CuO$_4$ plaquettes. In other
words, the ZN-model treats the electrons and holes in the $e-h$
pair substantially asymmetrically, that does not allow to
introduce the $S$- and $P$-like excitons.

The magnitude of the effective resonance two-particle transfer
integral $T_{AB}$ which determines the even-odd splitting is of
particular interest in exciton theory. In the real cuprate
situation the bare two-center exciton represents a system of two
neighboring electron CuO$_{4}^{7-}$ and hole CuO$_{4}^{5-}$
centers (given common oxygen). Thus, the resonance reaction
corresponds to inter-center transfer of {\it two holes}, or {\it
two electrons}.

The $S$-exciton is dipole-forbidden, in contrast to the
$P$-exciton, and corresponds to a so-called two-photon state.
However, these two excitons have a very strong dipole-coupling
with a large value of the $S$-$P$ transition dipole matrix
element. This points to a very important role played by this
doublet in nonlinear optics, in particular in two-photon
absorption and third-harmonic generation effects.
\cite{Ogasawara,THG}

The large dipole matrix element between excited $S$- and
$P$-excitons
\begin{equation}
\label{me} d = |\langle S|\hat{\bf d}|P\rangle |\approx 2eR_{CuCu}
,
\end{equation}
is important for the explanation of the nonlinear optical effects in
Sr$_2$CuO$_3$. \cite{Ogasawara,Kishida} The magnitude of this matrix
element yields a reliable estimate for the effective "length" of the
two-center CT exciton.

\subsection{Transition matrix elements for two-center excitons}

\subsubsection{General expressions}

In general, the expression for the EELS transition matrix element
for two-center excitons has a rather complicated form
\begin{widetext}
 $$
 \langle \Psi ^{g}_{ES}|e^{i{\bf k}{\bf r}}|\Psi_{GS}\rangle
 = \sin \alpha \cos \alpha  \left( -c_{ES}\langle
 \phi^{g}_{b_{1g}^{b}b_{1g}^{b}}|e^{i{\bf k}{\bf r}}| \phi
^{g}_{b_{1g}^{b}b_{1g}^{b}}\rangle
 +\sum _{\Gamma \Gamma ^{'}}
c_{\Gamma }^{*}a_{\Gamma ^{'}}\langle \phi
^{g}_{ES}(\Gamma)|e^{i{\bf k}{\bf r}}|\phi^{g}_{ES}(\Gamma
^{'})\rangle \right) ,
 $$
 \begin{equation}
\langle \Psi ^{u}_{ES}|e^{i{\bf k}{\bf r}}|\Psi_{GS}\rangle =
\sin\alpha  \sum _{\Gamma \Gamma ^{'}}b_{\gamma }^{*}
 a_{\gamma ^{'}}\langle \phi
^{u}_{ES}(\Gamma)|e^{i{\bf k}{\bf r}}|\Psi ^{g}_{ES}(\Gamma
^{'})\rangle . \label{GS-gu}
\end{equation}
\end{widetext}
It should be emphasized that it is the second term in the ground
state wave function (\ref{GS2}) which is particularly important
for the local contribution to the transition matrix elements. The
different transition matrix elements for the "two-plaquette"
Cu$_2$O$_7$ system can be easily reduced to "one-plaquette"
two-hole matrix elements. One has to use (\ref{eh1}) and
(\ref{gd},\ref{ud}) which give $\phi^{g,u}_{ES}(\Gamma)$ as
superpositions of the "one-plaquette" two-hole wave functions
$\Phi^{(2)}_{A,B}(\Gamma)$ centered at A and B. Then one finds for
the $x$-axis oriented Cu$_2$O$_7$ molecule
\begin{widetext}
$$
\langle \phi^{g}_{ES}(\Gamma)|e^{i{\bf k}{\bf r}}|\phi^{g}_{ES}
(\Gamma') \rangle = \cos\frac{k_{x}a}{2} \langle
\Phi_A^{(2)}(\Gamma)|e^{i{\bf k}{\bf
r}}|\Phi_A^{(2)}(\Gamma')\rangle ,
$$
\begin{equation}
\langle \phi_{ES}^{u}(\Gamma))|e^{i{\bf k}{\bf
r}}|\phi_{ES}^{g}(\Gamma')\rangle = -i\sin\frac{k_{x}a}{2}
\langle \Phi_A^{(2)}(\Gamma)|e^{i{\bf k}{\bf
r}}|\Phi_A^{(2)}(\Gamma')\rangle \label{gu}
\end{equation}
\end{widetext}
for $\Gamma$ and $\Gamma'$ of the same parity. Otherwise, the
right hand sides in (\ref{gu}) should be interchanged. At the
$\Gamma$-point ${\bf k}=0$ all these matrix elements are simply
reduced to appropriate overlap integrals. The $GS\rightarrow
g$-state transitions at this point are symmetry forbidden, while
the $GS\rightarrow u$-state transitions are allowed in the dipole
approximation (${\bf k}\rightarrow 0$). It is of particular
importance to note that in accordance with the relations
(\ref{gu}) the nonzero dipole response at the $\Gamma$-point is
derived only from either diagonal "one-plaquette" two-hole matrix
elements, or from non-diagonal ones with different parity of the
left and right hand side's functions.

\subsubsection{Some basic transition matrix elements for CT governed by
$\sigma$ bond}
  Below we list some nonzero
 matrix elements for the $x$-axis oriented two-center $S$- and $P$-excitons
 generated by the main $b_{1g}^{b}\rightarrow a_{1g},b_{1g},e_{u}$ CT
 transitions with the strongest $\sigma$ bonds:
 \begin{widetext}
One obtains for the local overlap contribution due to O
$2p_\sigma$ states for $\gamma ,\gamma ^{'}=a_{1g},b_{1g}$
 $$
\langle \phi ^{g}_{b_{1g}^{b}b_{1g}^{b}}|e^{i{\bf k}{\bf r}}| \phi
^{g}_{b_{1g}^{b}b_{1g}^{b}}\rangle =\rho _{GS}(O2pb_{1g}^{b})
 \cos\frac{k_{x}a}{2}(\cos\frac{k_{y}a}{2}+\cos\frac{k_{x}a}{2}),
 $$
 $$
\langle \phi ^{g}_{ES}(b_{1g}^{b}\gamma )|e^{i{\bf k}{\bf r}}|
 \phi ^{g}_{ES}(b_{1g}^{b}\gamma )\rangle =\frac{1}{2}
 (\rho _{ES}(O2p\gamma )+\rho _{GS}(O2pb_{1g}^{b}))
 \cos\frac{k_{x}a}{2}(\cos\frac{k_{y}a}{2}+\cos\frac{k_{x}a}{2}),
 $$
$$
\langle \phi ^{g}_{ES}(b_{1g}^{b}\gamma )|e^{i{\bf k}{\bf r}}|
 \phi ^{g}_{ES}(b_{1g}^{b}\gamma ^{'})\rangle =
 \frac{1}{2}c_{ES}^{*}(O2p\gamma )
 c_{ES}(O2p\gamma ^{'})
\cos\frac{k_{x}a}{2}(\cos\frac{k_{x}a}{2}-\cos\frac{k_{y}a}{2}),
\,(\gamma ^{'}\not=\gamma ),
 $$
$$
\langle \phi ^{u}_{ES}(b_{1g}^{b}\gamma )|e^{i{\bf k}{\bf r}}|
 \phi ^{g}_{ES}(b_{1g}^{b}\gamma )\rangle =-\frac{i}{2}
 (\rho _{ES}(O2p\gamma )+\rho _{GS}(O2pb_{1g}^{b}))
 \sin\frac{k_{x}a}{2}(\cos\frac{k_{y}a}{2}+\cos\frac{k_{x}a}{2}),
 $$
$$
\langle \phi ^{u}_{ES}(b_{1g}^{b}\gamma )|e^{i{\bf k}{\bf r}}|
 \phi ^{g}_{ES}(b_{1g}^{b}\gamma ^{'})\rangle =
 -\frac{i}{2}c_{ES}^{*}(O2p\gamma )
 c_{ES}(O2p\gamma ^{'})
\sin\frac{k_{x}a}{2}(\cos\frac{k_{x}a}{2}-\cos\frac{k_{y}a}{2}),
\,(\gamma ^{'}\not=\gamma );
$$
\begin{equation}
\langle \phi ^{g}_{ES}(b_{1g}^{b}e_{u}(\sigma))|e^{i{\bf k}{\bf
r}}|
 \phi ^{g}_{ES}(b_{1g}^{b}\gamma )\rangle =
 \frac{1}{2}c_{ES}^{*}(O2pe_{u}(\sigma) )c_{ES}(O2p\gamma )
\sin\frac{k_{x}a}{2}\sin\frac{k_{x}a}{2};\label{TME}
\end{equation}
$$
\langle \phi ^{u}_{ES}(b_{1g}^{b}e_{u}(\sigma))|e^{i{\bf k}{\bf
r}}|
 \phi ^{g}_{ES}(b_{1g}^{b}\gamma )\rangle =
 \frac{i}{2}c_{ES}^{*}(O2pe_{u}(\sigma) )c_{ES}(O2p\gamma )
\sin\frac{k_{x}a}{2}\cos\frac{k_{x}a}{2};
$$
 $$
\langle \phi ^{g}_{ES}(b_{1g}^{b}e_{u}(\sigma))|e^{i{\bf k}{\bf
r}}|
 \phi ^{g}_{ES}(b_{1g}^{b}e_{u}(\sigma))\rangle =
 \rho _{ES}(O2pe_{u}(\sigma))
\cos\frac{k_{x}a}{2}\cos\frac{k_{x}a}{2};
$$
$$
\langle \phi ^{u}_{ES}(b_{1g}^{b}e_{u}(\sigma))|e^{i{\bf k}{\bf
r}}|
 \phi ^{g}_{ES}(b_{1g}^{b}e_{u}(\sigma))\rangle =
 -i\rho _{ES}(O2pe_{u}(\sigma))
\sin\frac{k_{x}a}{2}\cos\frac{k_{x}a}{2};
$$
and for the local overlap contribution due to Cu $3d_{x^2 -y^2}$
states
 $$
 \langle \phi ^{g}_{b_{1g}^{b}b_{1g}^{b}}|e^{i{\bf k}{\bf r}}| \phi
^{g}_{b_{1g}^{b}b_{1g}^{b}}\rangle =2\rho _{GS}(Cu3db_{1g}^{b})
 \cos\frac{k_{x}a}{2}
 $$
$$
\langle \phi ^{g}_{ES}(b_{1g}^{b}\gamma )|e^{i{\bf k}{\bf r}}|
 \phi ^{g}_{ES}(b_{1g}^{b}\gamma )\rangle =
 (\rho _{ES}(Cu3d\gamma )+\rho _{GS}(Cu3db_{1g}^{b}))
\cos\frac{k_{x}a}{2},
 $$
$$
\langle \phi ^{u}_{ES}(b_{1g}^{b}\gamma )|e^{i{\bf k}{\bf r}}|
 \phi ^{g}_{ES}(b_{1g}^{b}\gamma )\rangle  =
 -i(\rho _{ES}(Cu3d\gamma )+\rho _{GS}(Cu3db_{1g}^{b}))
 \sin\frac{k_{x}a}{2},
  $$
 \begin{equation}
\langle \phi ^{g}_{ES}(b_{1g}^{b}e_{u}(\sigma))|e^{i{\bf k}{\bf
r}}|
 \phi ^{g}_{ES}(b_{1g}^{b}e_{u}(\sigma))\rangle =
 \rho _{GS}(Cu3db_{1g}^{b})\cos\frac{k_{x}a}{2},\label{TME1}
 \end{equation}
 $$
\langle \phi ^{u}_{ES}(b_{1g}^{b}e_{u}(\sigma))|e^{i{\bf k}{\bf
r}}|
 \phi ^{g}_{ES}(b_{1g}^{b}e_{u}(\sigma))\rangle =
 -i\rho _{GS}(Cu3db_{1g}^{b})\sin\frac{k_{x}a}{2},
 $$
 $$
\langle \phi ^{g}_{ES}(b_{1g}^{b}\gamma )|e^{i{\bf k}{\bf r}}|
 \phi ^{g}_{ES}(b_{1g}^{b}\gamma ^{'})\rangle =\langle \phi
^{u}_{ES}(b_{1g}^{b}\gamma )|e^{i{\bf k}{\bf r}}|
 \phi^{g}_{ES}(b_{1g}^{b}\gamma ^{'})\rangle = 0 , \,(\gamma
^{'}\not=\gamma ).
 $$
\end{widetext}
Here, $c_{GS,ES}(Cu3d\gamma ), c_{GS,ES}(O2p\gamma )$ and $\rho
=|c|^2$ are the
 probability amplitudes and hole densities
on copper (oxygen) molecular orbitals for ground, or excited states,
respectively.
 Making use of these expressions we can readily obtain  the complete set of
EELS transition
 matrix elements for dipole-allowed and -forbidden two-center excitons.
 Moreover, they can be used for the description of CT excitons in terms of
purely
 copper $3d\gamma$, or oxygen $2p\gamma$ one-hole states. Namely such
 an approach
 could be probably more reasonable in the case of strongly correlated
"one-plaquette"
 two-hole states.

 Formulas (\ref{GS-gu})-(\ref{TME1}) clearly show that the simple dipole
approximation for EELS transition
 matrix elements like that one used in the ZN-model \cite{Wang,Ng} cannot
satisfactorily describe the ${\bf k}$-dependence of
 the EELS intensities, and may be misleading.
 Interestingly, that the oxygen and copper contributions to the local overlap
 mechanism of EELS transitions have substantially different ${\bf
k}$-dependence.
 So, the O $2p$ contribution to both dipole-allowed and -forbidden
  transitions for $\gamma ,\gamma ^{'}=a_{1g},b_{1g}$ turn
  to zero at the $(\pi ,\pi)$  point, contrary to the copper
contribution. At this point the active
  oxygen contribution to the loss function is associated only with
dipole-forbidden
  $\Psi_{GS} \rightarrow \phi^{g}_{ES}(b_{1g}^{b}e_{u}(\sigma))$
  transitions. For the $[100]$ direction the oxygen contribution to the
  EELS intensity for the dipole-allowed transitions decreases twice as faster
  than the copper one.

  In other words, the EELS intensity at the $(\pi ,\pi)$  point
  reproduces in such a case the spectral distribution of the two-center
excitations generated
  by the $b_{1g}^{b}\rightarrow e_{u}$ CT transition. In contrast, the EELS
intensity
 near the $\Gamma$-point, where the $S$-exciton contribution turns into zero,
 reproduces the spectral distribution of the two-center
dipole-allowed excitations $g \rightarrow u$, generated   by the
$b_{1g}^{b}\rightarrow b_{1g}$ CT transitions both to
predominantly oxygen $(4 \rho ^{ES}(O2p\gamma ) > \rho
^{ES}(Cu3d\gamma ))$ and copper $(4 \rho ^{ES}(O2p\gamma )< \rho
^{ES}(Cu3d\gamma ))$ states, as well as the $b_{1g}^{b}\rightarrow
e_{u}(\pi),e_{u}(\sigma)$  CT transition to purely oxygen
$e_{u}(\pi),e_{u}(\sigma)$ states. In principle, this makes it
possible to examine the Cu $3d$ and O $2p$ partial composition of
the excitons by means  of the EELS intensity analysis, though it
should be noted that dipole-allowed two-center excitations could
interact both with each other and with the dipole-allowed
one-center excitations $b_{1g}^{b}\rightarrow
e_{u}(\pi),e_{u}(\sigma)$.

\subsubsection{Transition matrix elements for the $b_{1g}^2$-channel
of inter-center CT excitons}

For illustration, let us consider the $b_{1g}^2$-channel of
inter-center CT excitons. We have to take into account the
admixture of all three singlet wave functions $\Phi_i^{(2)}$ of
the ZR-singlet sector ((\ref{zr1})-(\ref{zr3})) to the ground
state wave function $\Psi_{GS}$. The corresponding amplitudes in
Eq.\ (\ref{GS2}) will be denoted by $a^{(i)}_{b_{1g}}$. So, the
$b_{1g}^2$-channel consists of three inter-center CT excitons with
energy separations of $\Delta E_{12}\approx 5.2$ eV and $\Delta
E_{13}\approx 6.7$ eV, respectively. If we neglect for a moment
the configuration interaction effects in the final states, the
relative intensities of these three excitons at the $\Gamma$-point
are given by:
\begin{equation}
 |a^{(1)}_{b_{1g}}|^2 :|a^{(2)}_{b_{1g}}|^2 :|a^{(3)}_{b_{1g}}|^2
 \; .
\end{equation}
More accurately, the intensity of the low-energy dipole-allowed
two-center exciton with the formation of the ZR-singlet
$b_{1g}^2;pd$ is determined by the matrix element:
$$ \langle
\phi^{u}_{ES}(b_{1g}^{2};pd)|e^{i{\bf k}{\bf r}}|
 \Psi_{GS}\rangle =-i \sin \alpha \sin \frac{k_{x}a}{2}\cdot
$$
$$
\left[ \langle d^{2}|e^{i{\bf k}{\bf r}}|d^{2}\rangle \left(
0.06a^{(1)}_{b_{1g}} - 0.24a^{(2)}_{b_{1g}} + 0.04a^{(3)}_{b_{1g}}
\right) \right.
$$
$$
+ \langle pd|e^{i{\bf k}{\bf r}}|pd \rangle \left(
0.90a^{(1)}_{b_{1g}} + 0.20a^{(2)}_{b_{1g}} - 0.22a^{(3)}_{b_{1g}}
\right)
$$
\begin{equation}
+ \left. \langle p^{2}|e^{i{\bf k}{\bf r}}|p^{2} \rangle \left(
0.04a^{(1)}_{b_{1g}} + 0.04a^{(2)}_{b_{1g}} + 0.18a^{(3)}_{b_{1g}}
\right) \right]
\end{equation}
where the one-plaquette matrix elements are:
$$
 \langle d^{2}|e^{i{\bf k}{\bf r}}|d^{2}\rangle =2 \; ,
$$
$$
 \langle pd|e^{i{\bf k}{\bf r}}|pd\rangle =
 1+\frac{1}{2}(\cos\frac{k_{x}a}{2}+\cos\frac{k_{y}a}{2}) \, ,
 $$
 \begin{equation}
 \langle p^{2}|e^{i{\bf k}{\bf r}}|p^{2}\rangle =
  (\cos\frac{k_{x}a}{2}+\cos\frac{k_{y}a}{2}) \; .
  \label{pd}
 \end{equation}
 Let us draw the attention to the different ${\bf k}$ dependence of the "diagonal"
 (the first term) and "non-diagonal" (the second and third terms) contributions
  to  the overall matrix element. When one moves from the $\Gamma$-point to $(\pi
,\pi)$, the absolute value of the former decreases, contrary to
the latter
 which grows from zero at the $\Gamma$-point to a non-zero value at the BZ
boundary.
 The final ${\bf k}$ behavior of the overall transition matrix element could be
 rather complicated depending on the relative signs and magnitudes of the
  $a^{(i)}_{b_{1g}}$ coefficients, or the admixture amplitudes of
  the different ZR-singlet-like states in the ground state wave function.
  These amplitudes depend mainly on the effective $b_{1g}^{b} \to b_{1g}$
   transfer integrals for the bare $b_{1g}^{b}$ hole between two neighboring
   plaquettes. The parameters suggest the following relationship
  \begin{equation}
  |a^{(3)}_{b_{1g}}| \,>\,|a^{(1)}_{b_{1g}}| \,>\,|a^{(2)}_{b_{1g}}|
  \, .
  \end{equation}
  In other words, the largest effective integrals will be expected for
  the transfer to predominantly oxygen $b_{1g}$-like hole states.
   In such a case, we cannot exclude the
  appearance of an intensity compensation point due to the competition of the
 "diagonal"   and "non-diagonal" terms in the matrix element. In any case, we may
  expect a rather   unusual ${\bf k}$ behavior of the transition matrix element
   which in turn   determines the behavior of the  EELS intensity for one of
   the main low-energy   contributions.  On the other hand,
   the large value of $a^{(3)}_{b_{1g}}$ leads to an higher
   intensity of the high energy CT exciton at the $\Gamma$-point in comparison to the
   low-energy one with an estimated energy separation between the two of
   $\approx 6.7$ eV. All this illustrates the particular importance
   of the correlation effects for two-hole configurations being the final
states
   for the inter-center CT transitions under consideration.  In addition, one
should
   note that a reasonable analysis of the $b_{1g}^{b} \to b_{1g}$ channel points
   to a rather wide spectral range of the intensive two-center CT transitions.

   If it were possible to neglect the effects of $d^{2}$- $pd$- and $p^{2}$-mixing
in the $b_{1g}^2$
   configurations, the dipole-allowed  $b_{1g}^2$-channel could be divided into
independent
  $d^{2}$- $pd$- and $p^{2}$-contributions with the corresponding EELS intensities
  $  I_{d^{2}}\propto |\langle d^{2}|e^{i{\bf k}{\bf r}}|d^{2}\rangle |^2 \, ,
   I_{pd}\propto |\langle pd|e^{i{\bf k}{\bf r}}|pd\rangle  |^2 \, ,
   I_{p^{2}}\propto |\langle p^{2}|e^{i{\bf k}{\bf r}}|p^{2}\rangle |^2 \,
  $
  (see the matrix elements from (\ref{pd})).

The intensities of dipole-forbidden ($g-g$) CT transitions in the
$b_{1g}^2$-channel turn into zero both at the $\Gamma $-point and at
the boundary of the Brillouin zone. To illustrate the ${\bf
k}$-dependence of the appropriate matrix elements, we restrict
ourselves below to the simplified model where only the ZR-singlet
contributes to the $b_{1g}^2$-channel
\begin{widetext}
\begin{equation}
    \langle \phi^{g}_{ES}(b_{1g}^{2};pd)|e^{i{\bf k}{\bf r}}|
 \Psi_{GS}\rangle =\frac{1}{2}\sin 2\alpha
 \left( \rho _{ES}(O2p)-\rho _{GS}(O2p) \right)\cos\frac{k_{x}a}{2}
 \left[ \frac{1}{2}(\cos\frac{k_{x}a}{2}+\cos\frac{k_{y}a}{2})-1 \right] \; .
\end{equation}
\end{widetext}
 It should be noted that for the [11] and [10] directions the matrix
 element reaches  its maximum at
 $k=\frac{2}{3}k_{max}$. This value depends on the actual difference in
 the $b_{1g}$ hole density distribution of ground and excited states.

\subsubsection{Transition matrix elements for the $b_{1g}^{b}a_{1g}$-channel
of inter-center CT excitons }

Taking into account the correlation and configuration interaction
effects there appear three  ${}^{1}B_{1g}$ terms with a
$b_{1g}a_{1g}$-like configuration which form the final states for
the $b_{1g}^{b}a_{1g}$-channel of the two-center CT transitions.
This channel is only governed by the oxygen contribution, due to the
orthogonality of the Cu $3db_{1g}$ and Cu $3da_{1g}$ orbitals. A
shorthand analysis of the ${\bf k}$ dependence for the different
matrix elements listed above in Eqs.\ (\ref{TME})-(\ref{TME1}) shows
that the $b_{1g}^{b}a_{1g}$-channel is "silent" both at the
$\Gamma$-point and all along the $(\pi ,\pi)$ direction. Along the
$(\pi ,0)$ direction we deal with an unusual behavior  of the
appropriate $P$($A_u$) and $S$($A_g$) excitonic modes. Going to the
BZ boundary, the intensity of the former mode increases while that
of the latter one turns into zero both at $\Gamma$  and at $(\pi
,0)$ with nonzero value in between. So, one may conclude that the
$b_{1g}^{b}a_{1g}$-channel could be revealed only along the $(\pi
,0)$ direction outside the $\Gamma$-point.

 \subsubsection{Transition matrix elements for the $b_{1g}^{b}e_{u}$-channel
 of inter-center CT excitons }

 Like the $b_{1g}^{b}a_{1g}$-channel the $b_{1g}^{b}e_{u}$-channel
 is only governed by the oxygen contribution.
 As it was mentioned above, we have two types of purely oxygen $e_{u}$
orbitals:
 $e_{u}(\sigma)$ and $e_{u}(\pi)$ with $\sigma$ and $\pi$ directions of the O $2p$
 lobes, respectively. These orbitals hybridize with each other owing
 to the O $2p$-O $2p$
 transfer. As a result we obtain four ${}^{1}E_u$ terms for the
$b_{1g}e_u$-like
 two-hole configurations ($b_{1g}^{b}e_{u}(\pi) ,b_{1g}^{b}e_{u}(\sigma) ,
  b_{1g}^{a}e_{u}(\pi) ,b_{1g}^{a}e_{u}(\sigma$))
  which form the final state of the
 $b_{1g}^{b}e_{u}$-channel of the two-center CT transitions.

 Non-zero transition matrix elements in Eqs.\ (\ref{TME})-(\ref{TME1})
 are only obtained for the $b_{1g}^{b}e_{u}(\sigma)$-channel.  Therefore, the
resulting
 magnitude of the transition matrix elements would be firstly determined by the
 $e_{u}(\sigma)$ weight in the $b_{1g}e_u$ configuration for the final
${}^{1}E_u$ term. On the other hand, the main contribution to
transition matrix elements is determined by the
 purely oxygen $b_{1g}$ component of the final states. Both these conclusions
  are of particular importance for the relative intensities in the
  $b_{1g}^{b}e_{u}$-channel. The doublet of transitions with final antibonding
  $b_{1g}^{a}$ state is more intensive than the similar doublet  with the final bonding
  $b_{1g}^{b}$ state: its relative intensity could be estimated to be
  $|\tan\alpha _{b_{1g}}|^{2}\approx 1.3$. For either doublet the strongest
   intensity is predicted for
 transitions with final   $e_{u}(\sigma)$ state with estimated relative intensity
 to be $|\tan\alpha _{e_{u}}|^{2}>1$.

 Different
theoretical estimations (see above) point to the predominantly
$b_{1g}e_{u}(\pi)$ structure of the lowest in energy
 ${}^{1}E_u$ term of the two-hole CuO$_{4}^{5-}$ center, whose separation from
the ground state ZR singlet is only about $1.5\div 2.0$ eV as
observed in photoemission experiments \cite{Pothuizen} for
Sr$_2$CuO$_2$Cl$_2$.
 Thus, one should expect rather moderate intensities for
the appropriate excitonic mode. On the other hand, one might
naturally expect a rather strong high-energy transition to the
predominantly $b_{1g}^{a}e_{u}(\sigma)$ state. Overall, the ${\bf
k}$ dependence of the transition matrix elements for the
 $b_{1g}^{b}e_{u}$-channel is rather typical for purely oxygen contributions:
the dipole-allowed $P$ modes gradually loose their intensity up to
zero going
 from the  $\Gamma$-point  to the BZ boundary both in $(\pi ,\pi)$ and $(\pi ,0)$
directions, contrary to the dipole-forbidden  $S$ modes, whose
intensity rises by approaching the BZ boundary.

 It should be noted that the relative energy position of the four CT
excitations
 in the $b_{1g}^{b}e_{u}$-channel is determined by the $e_{u}(\pi) -e_{u}(\sigma)$
 and the $b_{1g}^{b}-b_{1g}^{a}$ separations, which are of the order $5.0$ and
  $6.0$ eV, respectively, as predicted by the above quantum-chemical cluster
   calculations.

\section{Dynamics and dispersion of small excitons}

\subsection{Rotational and translational motion of  two-center excitons}

\begin{figure}[h]
\includegraphics[width=8.5cm,angle=0]{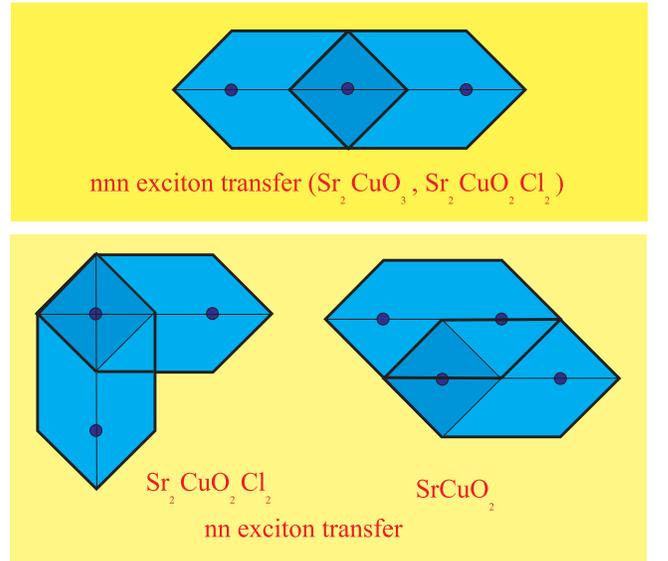}
\caption{Different types of next-nearest (nn) neighbor or second
nearest neighbor transfer, as well as next-next nearest (nnn)
neighbor or 3rd nearest neighbor transfer of two-center excitons.}
\label{fig4}
\end{figure}

In contrast to one-center Frenkel excitons, the motion of its
two-center counterparts  is more complicated. We addressed above the
internal electron-hole  motion resulting in $S-P$ splitting. Now, we
consider the rotational and translational motion of two-center
excitons. The elongated structure of the two-center exciton results
in a set of both rotational and translational modes, depending on
the dimension of the lattice of CuO$_4$ centers. Such modes for the
2D lattice were addressed in the frames of the simple ZN-model:
\cite{Wang,Ng} these are rotations of the hole (electron) around the
electron (hole) by $90^o$ and $180^o$, and axial translations (see
Fig.\ \ref{fig4}).
Strictly speaking, Zhang and Ng in their simple model considered
only the $90^o$ and $180^o$ rotation around  the electron center
with the respective transfer integrals $t_1$ and $t_4$, and the
axial translation accompanied by a $90^o$ rotation around  the
electron center with transfer integral $t_2$.

It is interesting to note that the $90^o$- and $180^o$-rotation of
the hole around the electron corresponds to the 2nd nearest
neighbor and 3rd nearest neighbor hopping of the hole
CuO$_{4}^{5-}$ center which is coupled to a corresponding motion
of the electron CuO$_{4}^{7-}$ center. This problem is closely
related to the well-known problem of the ZR-singlet being the
ground state of the CuO$_{4}^{5-}$ center, moving in the lattice
formed by the CuO$_{4}^{6-}$ centers. \cite{Roland}

\subsubsection{Exciton transfer integrals}

Let us consider a three center system ABC and let us compare the
collinear $180^o$ geometry with the rectangular $90^o$ one. The
exciton transfer integrals between two-center excitons centered at
AB or BC, respectively, can be expressed through matrix elements
of the effective Hamiltonian $H_{eff}$ that incorporates potential
and kinetic energy contributions:
$$
\frac{1}{2}\langle \Phi_{A}^{(0)}\Phi_{B}^{(2)}\pm
\Phi_{A}^{(2)}\Phi_{B}^{(0)}| {\hat H}_{eff}| \Phi_{B}^{(0)}
\Phi_{C}^{(2)} \pm \Phi_{B}^{(2)}\Phi_{C}^{(0)}\rangle =
$$
\begin{equation}
\pm \frac{1}{2} \left( T_{e}^{(2,3)} + T_{h}^{(2,3)} \right) +
\ldots \label{dia}
\end{equation}
for diagonal $SS$ and $PP$ transfer, and
$$
\frac{1}{2}\langle \Phi_{A}^{(0)}\Phi_{B}^{(2)} \pm
\Phi_{A}^{(2)}\Phi_{B}^{(0)} | {\hat H}_{eff}| \Phi_{B}^{(0)}
\Phi_{C}^{(2)} \mp \Phi_{B}^{(2)} \Phi_{C}^{(0)}\rangle =
$$
\begin{equation}
\pm \frac{1}{2} \left( T_{e}^{(2,3)} - T_{h}^{(2,3)} \right) +
\ldots \label{ndia}
\end{equation}
for off-diagonal $SP$ transfer. Here, $T_{e,h}^{(2,3)}$ are the
electron (hole) transfer integrals. One has to distinguish the
collinear geometry with an exciton transfer to 3rd neighbors
$T_{e,h}^{(3)}$ separated by $\approx 4$\AA, from the rectangular
geometry with a transfer to 2nd nearest neighbors
$T_{(e,h)}^{(2)}$ separated by $\approx \frac{4}{\sqrt{2}}$\AA. In
(\ref{dia}),(\ref{ndia}) we neglected higher order terms which are
less important.

According to Eqs.\ (\ref{dia}),(\ref{ndia}) we introduce a set of
transfer parameters to describe the exciton dynamics:
$$
T_{S} \approx  T_{P} \approx \frac{1}{2}(T_{e}^{(3)}+T_{h}^{(3)})
\; ,
$$
$$
T_{SP}\approx  \frac{1}{2}(T_{e}^{(3)}-T_{h}^{(3)}) \; ,
$$
for the collinear exciton motion, and
$$
T_{S}^{xy}\approx T_{P}^{xy} \approx
 \frac{1}{2}(T_{e}^{(2)}+T_{h}^{(2)}) \; ,
$$
$$
T_{SP}^{xy}\approx  \frac{1}{2}(T_{e}^{(2)}-T_{h}^{(2)}) \; ,
$$
corresponding to a $90^o$ rotation of the exciton.

All these parameters have a rather clear physical sense. Moreover,
the electron (hole) transfer integrals for collinear exciton
transfer $T_{e,h}^{(3)}$ are believed to be smaller  than
$T_{e,h}^{(2)}$ integrals for rectangular transfer. In other words,
the two-center excitons prefer to move "crab-like", rather than in
the usual collinear mode. This implies a large difference for the
excitonic dispersion in [10] and [11] directions. The electronic
wave function in the exciton (contrary to the hole one) has a
dominant Cu $3d$ nature that implies a smaller magnitude of the
$T_{e}^{(2,3)}$ parameters compared to the $T_{h}^{(2,3)}$ ones.

As it was shown by Zhang and Ng, \cite{Wang,Ng} the singlet
two-center $S,P$ exciton can move through the antiferromagnetic
lattice rather freely in contrast with the single-hole motion.
Nevertheless, their values for the second order integrals $t_3
\approx 0.85$ eV and $t_4 \approx 0.65$ eV as derived from an
analysis of the experimental EELS data in terms of the simple
ZN-model, might be considerably overestimated. As will be seen in
the more realistic analysis of EELS data presented below, there
are several channels which contribute to the exciton dispersion,
such that the numerical values of each of the exciton transfer
parameters might be substantially smaller than the effective
transfer integrals $t_{1\div 4}$ of the simple ZN-model.

Up to now, we addressed the two-center exciton dynamics in
corner-shared systems like Sr$_2$CuO$_3$ or Sr$_2$CuO$_2$Cl$_2$.
However, in 1D systems such as SrCuO$_2$ there is an additional
type of nearest-neighbor transfer of two-center excitons localized
on the Cu$_2$O$_7$ cluster (as presented in Fig.\ \ref{fig4}).

\subsubsection{Tetragonal symmetry and two-center excitons
in the 2D CuO$_2$ layer}

The tetragonal symmetry of the 2D CuO$_2$ layer leads to certain
modifications of the results obtained above for excitons in the
simple Cu$_2$O$_7$ cluster. There, one had to distinguish $S$- and
$P$-like two-center excitons centered at the common oxygen site, in
analogy to the symmetry properties of O $2s$ and O $2p\sigma$
orbitals, respectively. In contrast to that, two-center excitons in
the 2D CuO$_2$ layer could be readily classified with regard to
irreducible representations of the tetragonal $D_{4h}$ group. Thus,
we can form a set of "large orbital" excitons with $A_{1g}$,
$B_{1g}$ and $E_{u}$ symmetry (or equivalently with $S$, $D$ and
$P_{x,y}$ symmetry, respectively) rotating around the Cu site (with
Cu$_{5}$O$_{16}$ geometry). Each two-center exciton classified by
the two-hole state $\Gamma$ of the CuO$_4^{5-}$ center, give rise to
one set of 4 excitons.

The symmetry of the excited CT states interacting with the ground
state has a crucial influence on the transition matrix elements. So,
given the $b_{1g}$ symmetry of the ground hole state in the isolated
CuO$_4$ plaquette, it is the "large orbital" charge transfer state
with the same $B_{1g}$ symmetry generated by $S_{x,y}$ two-center
transitions that can be mixed with the ground state of the large
Cu$_{5}$O$_{16}$ cluster. In terms of the 4 small $S_{x,y},P_{x,y}$
excitons in the 2D elementary cell this corresponds to the mode
$S_{-}=\frac{1}{\sqrt{2}}(S_{x}-S_{y})$. Naturally, the large
$E_{u}$ excitons generated by small $P_{x,y}$ excitons are
dipole-allowed, while both $A_{1g}$ and $B_{1g}$ excitons are
dipole-forbidden. In terms of the elementary cell modes the
dipole-allowed ones are the $P_{x,y}$, or
$P_{\pm}=\frac{1}{\sqrt{2}}(P_{x}\pm P_{y})$ modes.

The mixing of the $S_{-}$ mode to the ground state in the 2D system
defines the intensity of the two-center exciton in the CuO$_2$
layer, similarly to the simple $S_{x}$ mode in 1D systems. The
transition matrix elements for dipole-allowed excitons in 2D systems
is expressed through dipole matrix elements $\langle S_{-}|{\hat
d}|P_{x,y}\rangle$ instead of $\langle S_{x,y}|{\hat
d}|P_{x,y}\rangle$ in 1D systems; in other words they would be a
factor $\sqrt{2}$ smaller. Thus, the expected intensity of  the
dipole-allowed two-center CT transitions in 2D system is factor $2$
smaller than in 1D systems.

\subsection{The dynamics of two-center excitons in 1D and 2D cuprates}

In general, we will consider the  two-center $\Gamma$-excitons,
whose dynamics in frame of the Heitler-London approximation
\cite{Davydov} could be described by an effective one-particle
excitonic Hamiltonian with standard form
\begin{equation}
\hat H_{exc}=\sum _{\Gamma _{1}\Gamma _{2}{\bf R}_{1}{\bf R}_{2}}
\hat B^{\dagger}_{\Gamma _{1}}({\bf R}_{1}) T_{\Gamma _{1}\Gamma
_{2}}({\bf R}_{1}-{\bf R}_{2}) \hat B_{\Gamma _{2}} ({\bf R}_{2})
\end{equation}
in a site representation with $\hat B^{\dagger}_{\Gamma _{1}}({\bf
R}_{1})/ \hat B_{\Gamma _{2}} ({\bf R}_{2})$ being the excitonic
creation/annihilation operators, or
\begin{equation}
\hat H_{exc}=\sum _{\Gamma _{1}\Gamma _{2}{\bf k}} \hat
B^{\dagger}_{\Gamma _{1}}({\bf k})T_{\Gamma _{1}:\Gamma _{2}}
 ({\bf k})\hat B_{\Gamma _{2}} ({\bf k})
\end{equation}
in ${\bf k}$ representation. Here the $\Gamma _{1,2}$ indices label
different $S$- or $P$-excitons. One should note that instead of the
$S$-$P$ manifold we could consider also the dynamics of large
orbital excitons with a certain point group symmetry at the
$\Gamma$-point.

\subsubsection{The dynamics of isolated $SP$ excitonic doublet in 1D cuprates}

We consider the 1D cuprate as a linear chain of corner shared
CuO$_4$ centers directed along the $x$-axis. The $T({\bf k})$ matrix
for an isolated doublet of $S_x =S$ and
 $P_x =P$ excitons in the 1D
cuprate can be written in a rather simple form \cite{Davydov}
\begin{equation}
T({\bf k})=\pmatrix{E_{S}+2T_{S}\cos k_{x}&-2iT_{SP}\sin k_{x}\cr
2iT_{SP}\sin k_{x}&E_{P}+2T_{P}\cos k_{x}\cr}.
\end{equation}
The dispersion of the two-center exciton in 1D system is governed by
the nnn exciton transfer. As it was emphasized above, both
$(T_{S}-T_{P})$ and $T_{SP}$ are believed
 to be small that results in the near-degeneracy of the $S$ and $P$ modes
 with rather trivial dispersion.

The intensity of the dipole-allowed $P$ exciton is expected to
decrease sharply with the momentum, while that of the
dipole-forbidden $S$-exciton exhibits a more complicated behavior
with zeros at the $\Gamma$-point and the BZ boundary, and the
maximum midway between the center and the boundary of the zone at
$k=2/3 k_{max}$. Overall, one should note that all isolated $SP$
excitonic doublets can manifest a rather small dispersion, however,
the ${\bf k}$-dependence of the EELS intensities could be extremely
large, and they differ significantly for different transitions.
Naturally, in practice one should account for inter-exciton
coupling. It is interesting to note that the $S$- and $P$-excitons
are coupled by dipole transition with large matrix element, which
along with the $SP$ degeneracy makes the $SP$ excitonic doublet in
1D cuprates a very promising system for  nonlinear optical
devices.\cite{Ogasawara,THG}

\subsubsection{Dynamics of isolated $SP$ excitonic quartet in 2D cuprates}

In the 2D case of an ideal CuO$_2$ layer we deal with two types of
$x$- ($S_{x},P_{x}$) and $y$- ($S_{y},P_{y}$) oriented $S,P$
excitons in every elementary cell. The $T({\bf k})$ matrix for an
isolated quartet of $S_{x,y}$ and $P_{x,y}$ excitons in 2D cuprates
can be written  by a slight modification of the simple ZN-form
\cite{Ng}
\begin{widetext}
 \begin{equation}
T({\bf k})= \pmatrix{E_{S}+2T_{S}\cos k_{x}&-2iT_{SP}\sin k_{x}&
T_{S}^{xy}(1+a(k_{x},k_{y}))& T_{SP}^{xy}(1+b(k_{x},k_{y}))\cr
2iT_{SP}\sin k_{x}&E_{P}+2T_{P}\cos k_{x}&
T_{SP}^{xy}(1-b(k_{x},k_{y}))& T_{P}^{xy}(1-a(k_{x},k_{y}))\cr
T_{S}^{xy}(1+a^{*}(k_{x},k_{y}))& T_{SP}^{xy}(1-b^{*}(k_{x},k_{y}))&
E_{S}+2T_{S}\cos k_{y}&-2iT_{SP}\sin k_{y}\cr
T_{SP}^{xy}(1+b^{*}(k_{x},k_{y}))& T_{P}^{xy}(1-a^{*}(k_{x},k_{y}))&
2iT_{SP}\sin k_{y}&E_{P}+2T_{P}\cos k_{y}\cr}, \label{matrix}
\end{equation}
\end{widetext}
where $a(k_{x},k_{y})=e^{ik_{x}}+e^{-ik_{y}}$,
$b(k_{x},k_{y})=e^{ik_{x}}-e^{-ik_{y}}$. Here, the two $2\times 2$
blocks on the diagonal are related to $S_x ,P_x$,  $S_y ,P_y$
excitons, respectively;  off-diagonal blocks describe its coupling.
For the [11] direction the exciton transfer matrix  breaks up into
two similar $2\times 2$ blocks
\begin{widetext}
\begin{equation}
T_{\pm}({\bf k})=\pmatrix{E_{S}+2T_{S}\cos k_{x}\pm
T_{SS}^{xy}(1+2\cos k_{x})&\mp (2iT_{SP}\sin
k_{x}-T_{SP}^{xy}(1+2i\sin k_{x}))\cr \mp (-2iT_{SP}\sin
k_{x}-T_{SP}^{xy}(1-2i\sin k_{x}))&E_{P}+2T_{P}\cos k_{x}\pm
T_{PP}^{xy}(1-2\cos k_{x})\cr}.
\end{equation}
\end{widetext}
with basis vectors
\begin{equation}
|S_{\pm}\rangle =\frac{1}{\sqrt{2}}(|S_{x}\rangle \pm|S_{y}\rangle
); \quad |P_{\pm}\rangle =\frac{1}{\sqrt{2}}(|P_{x}\rangle
\pm|P_{y}\rangle ),
\end{equation}
respectively. In other words, we have two hybrid $S_{+}P_{+}$, and
two hybrid $S_{-}P_{-}$ modes. Generally speaking, all these modes
have different energies at ${\bf k}=0$ (the $\Gamma$-point). Both
hybrid $S_{-}P_{-}$ excitons involve the dipole-allowed
 (for ${\bf k}\parallel [11]$) $P_{-}$ mode and
can manifest itself in optical spectra.

In a nearest-neighbor approximation for the exciton transfer, when
$T_{S}=T_{P}=T_{SP}=0$, $E_{S}=E_{P}$, $T_{SS}^{xy}=T_{PP}^{xy}=T;
T_{SP}^{xy}=T_1$ we obtain four  modes with energies:
$$
E_{1}^{\pm} = E\pm T+[4\cos
^{2}k_{x}(T^{2}-T_{1}^{2})+5T_{1}^{2}]^{\frac{1}{2}};
$$
$$
E_{2}^{\pm} = E\pm T-[4\cos
^{2}k_{x}(T^{2}-T_{1}^{2})+5T_{1}^{2}]^{\frac{1}{2}}
$$
and eigenvectors:
$$
|(SP)^{(1)}_{\pm}\rangle = e^{i\phi}\cos \alpha |S_{\pm}\rangle +
e^{-i\phi}\sin \alpha |P_{\pm}\rangle ;
$$
$$
|(SP)^{(2)}_{\pm}\rangle = e^{i\phi}\sin \alpha |S_{\pm}\rangle
-e^{-i\phi} \cos \alpha |P_{\pm}\rangle ,
$$
where
$$
\tan 2\alpha = \frac{T_{1}[1+4\sin ^{2}k_{x}]^{\frac{1}{2}}}{2T\cos
k_{x}}; \quad \tan \phi =2\sin k_{x}.
$$
It should be noted that the complex matrix elements in $T_{\pm}$
given $k_{x}=k_{y}\not= 0$ evidence the  appearance of "{\it
twisting}" excitonic modes. If we can neglect the nnn electron
transfer in comparison with the hole transfer, then we obtain
$T=T_1$ and unexpectedly four dispersionless modes with energies:
$$
E_{1}^{\pm} = E \pm T + \sqrt{5} T; \quad E_{2}^{\pm} = E \pm T -
\sqrt{5} T,
$$
though the eigenvectors continue to depend on $k_x$ value.

If the electron and hole transfer  integrals have the same magnitude
($T_{e}^{(2)}=T_{h}^{(2)}$), then $T_{1}=0$ and we obtain pure
$|S_{\pm},P_{\pm}\rangle$ modes with a rather conventional
dispersion:
$$
E_{S}^{\pm} = E\pm T(1+ 2\cos k_{x}); \quad E_{P}^{\pm} = E\pm T(1-
2\cos k_{x}).
$$
Finally, we address the situation with $T_{e}^{(2)}=-T_{h}^{(2)}$,
when $T=0$, and we have two doubly degenerate twisting modes
$$
E_{1}^{\pm} = E+T_{1}[1+4\sin ^{2}k_{x}]^{\frac{1}{2}}; \quad
E_{2}^{\pm} = E-T_{1}[1+4\sin ^{2}k_{x}]^{\frac{1}{2}}\,
$$
with equal weight of $S$ and $P$ modes in eigenvectors.

In frame of the nn-model for the exciton transfer along the [10]
direction we address the only example with zero $SP$-mixing when the
exciton dynamics
 is governed by the only nonzero parameter $T$. Due to the "crab-like" motion
we come to
 four
 twisting modes with $S_{x}-S_{y}$ and $P_{x}-P_{y}$ mixing and energy
 $$
 E_{S_{\pm}}= E_{S} \pm T[(2+\cos k_{x})^{2}+\sin ^{2}k_{x})]^{\frac{1}{2}};
\quad
 E_{P_{\pm}}= E_{P} \pm |T|.
 $$
Here, the dipole-allowed $P_{\pm}$ modes appear to be
dispersionless. Interestingly to note, that instead of one expected
dipole-allowed mode ($P_x$, or $P_y$) along this direction, we
obtain $two$ dipole-allowed modes due to $P_{x}-P_{y}$ mixing.
Naturally, above we addressed rather simple "toy" models, however,
they allow to clearly demonstrate different probable scenarios for
the exciton dynamics and its manifestation in optical and EELS
spectra.

\subsubsection{The interaction of two-center excitons in 2D cuprates: The
interference effects}

Any two excitons of the same symmetry interact with each other.
The structure of the interaction matrix for two $S_{x,y},P_{x,y}$
quartets
 is similar to (\ref{matrix}).
The inter-exciton coupling effects its dispersion, and the
intensity. The latter is of particular importance for relatively
weak transitions in neighborhood of the strong ones. Depending on
the phase of the coupling matrix element, the intensity of the
definite CT transition can either increase, or decrease due to its
interaction with neighbors up to overall suppression. We would like
to argue that the particular structure of the interaction matrix
elements for the dipole-allowed $P_{x,y}$ excitons makes the
appearance of the intensity compensation points a general rule
rather than an exception. Indeed, the $P_{x,y}-P_{x,y}^{'}$ coupling
is governed by matrix elements $\langle P_{x,y}|T({\bf
k})|P_{x,y}^{'}\rangle \propto \cos k_{x,y}$ that change sign midway
between $\Gamma$-point and the BZ boundary. The
$P_{x,y}-P_{y,x}^{'}$ coupling along the [11] direction is governed
by matrix elements $\langle P_{x,y}|T({\bf k})|P_{y,x}^{'}\rangle
\propto (1- 2\cos k_{x,y})$
 that
as well change sign at $k_{x,y}=\frac{1}{3}k_{max}$.

To illustrate inter-exciton coupling effects we  address two
dipole-allowed two-center excitons, say $b_{1g}^2$ and $b_{1g}e_{u}$
type along the [11] direction with the only nonzero $P-P^{'}$
coupling. For simplicity, we assume a conventional bare energy
dispersion for both  $P$ modes:
$$
E(k) = E_{P}(0) + T_{P}\cos \frac{k}{\sqrt{2}}
$$
with $E_{P}(0)= 3.1, T_{P^{'}}= 0.4$ eV and $E_{P^{'}}(0)= 4.2,
T_{P}= 0.2$ eV, respectively, and conventional EELS intensity
dispersion:
$$
I_{P}(k) = I_{P}(0)\frac{2\sin
^{2}\frac{k}{2\sqrt{2}}}{k^{2}}(1+\cos \frac{k}{2\sqrt{2}})^2;
$$
$$
I_{P^{'}}(k) = I_{P^{'}}(0)\frac{\sin
^{2}\frac{k}{2\sqrt{2}}}{2k^{2}}
$$
typical for the $b_{1g}^2$ and $b_{1g}e_{u}$-channels, respectively,
with intensity ratio like: $I_{P}(0):I_{P^{'}}(0)= 1:10$. In other
words, we have neighboring weak low-energy and strong high-energy
excitons. The interaction matrix element is assumed to be
$$
\langle P|T(k)|P^{'}\rangle =T_{PP^{'}}(1- \cos \frac{k}{\sqrt{2}})
$$
to provide the non-interacting excitons at the $\Gamma$-point. It is
easy to see that the noticeable interaction effects manifest itself
only in the range $(0.5\div 1.0)k_{max}$. The sign and magnitude of
$T_{PP^{'}}$ were chosen
 to provide the intensity compensation point near $k=0.7k_{max}$:
  $|T_{PP^{'}}|= 0.3$ eV.
\begin{figure}[h]
\includegraphics[width=8.5cm,angle=0]{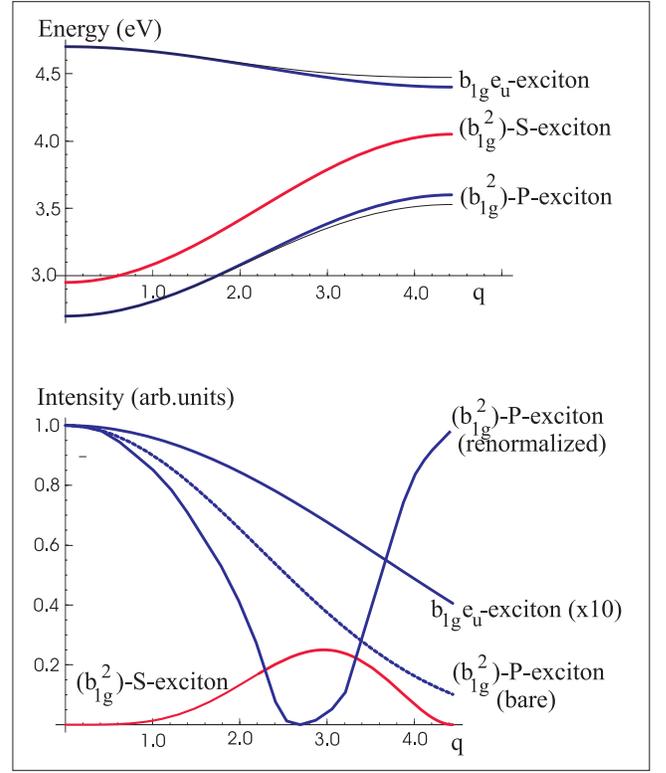}
\caption{Model dispersion curves for the energy (upper panel) and
 the EELS intensity (lower panel) for two neighboring $P$ excitons.
 For illustration we present the EELS intensity dispersion for isolated $S$
exciton. The momentum dependence of the EELS intensity for
$b_{1g}^2$-P-exciton demonstrates the destructive interference
effect with a full "bleaching" near certain critical q-value. See
text for details.} \label{fig5}
\end{figure}
The Fig.\ \ref{fig5} presents the model dispersion curves for the
energy (upper panel)
 together with those for the EELS
intensity (lower panel) for two $P$ excitons. We see that the
interaction results in a rather weak renormalization of the bare
energy dispersion for both excitons. However, the relatively weak
interaction leads to a cruciable renormalization of the intensity
dispersion for weak low-energy exciton with compensation point near
$q=0.7q_{max}$ and blazing up near BZ boundary. The appropriate
renormalization for the intensity of the strong high-energy exciton
is relatively small. Interestingly to notice that the compensation
effect stems from the destructive interference of the two excitons
and takes place in a rather narrow momentum range. In experimental
spectra this compensation can look like an effect of the "lost"
band.

 So, we see that the isolated $SP$ excitonic
quartets can manifest a rather moderate dispersion, however, the
${\bf k}$-dependence of the EELS intensities is always very strong
and  differs significantly for different transitions. Overall, the
model curves in Fig.\ \ref{fig5}
 clearly demonstrate an extremely important role played by the EELS
intensity dispersion. This "matrix element effect", or interference
effect is likely to complicate the theoretical treatment of the
experimental  EELS and RIXS spectra.

\subsubsection{Two-center excitons in 2D systems as compared with 1D systems}
Concluding this subsection we would like to make a short comparative
analysis of the main properties of two-center excitons in 2D and 1D
systems with corner-shared CuO$_4$ plaquettes:

1. In 1D system we deal only with nnn exciton transfer, while in 2D
system both nnn and nn transfer exists.

2. At the $\Gamma$-point we have in common 2 dipole-allowed
transitions associated with  one $SP$ excitonic manifold for 2D
system contrary to single transition in 1D system.

3. Integral intensity of dipole-allowed transitions associated with
one $SP$ excitonic manifold for 2D system two times smaller than in
1D system.

4.  The $S-P$ splitting in 1D system is expected to be substantially
smaller than in 2D systems. In addition, the $S-P$ dipole matrix
elements in 1D system are expected to be as a minimum two times
bigger than in 2D systems. Both these quantities essentially
determine the magnitude of the third order nonlinear susceptibility
$\chi ^{(3)}$ and a potential of the system as an optoelectronic material
with high performance.\cite{Ogasawara} Hence, 1D systems should be
more  effective nonlinear systems as compared with 2D counterparts.
 Experimental data for 1D Sr$_2$CuO$_3$ and 2D Sr$_2$CuO$_2$Cl$_2$,
and exact diagonalization technique on small clusters confirm this
conclusion.\cite{Ogasawara,nonlinear}

\section{CT excitons in 0D, 1D, and 2D insulating cuprates probed by  EELS and
optical spectroscopy}

By addressing the nature of the electron-hole excitations  we should
emphasize that a comparative analysis  of the optical and EELS
spectra in cuprates with different dimensionality of  the CuO$_4$
network  is extremely interesting and informative in many respects.
In particular, the considerable general interest in comparing the
different physical properties of the  0D, 1D and 2D members of the
growing cuprate family is warmed up by challenging problems such as
the low-dimensional aspect of electronic structure and high-$T_c$
superconductivity.

The 0D cuprates imply a system with well isolated or weakly coupled
CuO$_4$ plaquettes. Such a situation seems to be realized in
CuB$_2$O$_4$ which belongs to the class of noncentrosymmetric
magnetically ordered materials for which the unusual coexistence of
weak Dzyaloshinskii-Moriya type ferromagnetism and inhomogeneous
(incommensurate) magnetic ordering attracts a lot of attention.
CuB$_2$O$_4$ crystallizes in the tetragonal space group
$I\overline{4}2d$.\cite{CuBO} The Cu$^{2+}$ ions at 4b sites are
surrounded by four oxygen atoms in planar quadratic coordination so
that the local symmetry is $\overline{4}$. The Cu$^{2+}$ ions at 8d
sites occupy distorted octahedral positions with an exceptionally
large separation of 3.069 \AA \ from the two apical Cu$^{2+}$ ions.
Its local symmetry is 2.

Many insulating cuprates represent text-book low-dimensional model
systems. The 1D corner-shared ordering of the CuO$_4$ plaquettes is
realized  in insulating Sr$_2$CuO$_3$ to be the best known model
system for a spin-$1/2$  antiferromagnetic Heisenberg chain and a
very promising material for nonlinear  optical devices.
\cite{Ogasawara}

One of the best representatives for an insulating layered 2D copper
oxide might be the oxychloride Sr$_2$CuO$_2$Cl$_2$, which is
iso-structural to the famous 214 system La$_2$CuO$_4$ but contrary
to the latter has a single type both of copper and oxygen sites. The
compound Sr$_2$CuO$_2$Cl$_2$ is one of the most popular model system
for the insulating phase of the high-$T_c$ cuprates and 2D
spin-$1/2$ Heisenberg antiferromagnets,  and intensively studied
both experimentally and theoretically. In this tetragonal
antiferromagnet ($T_N \approx 250$ K) with nearly ideal CuO$_2$
planes there are chlorine atoms instead of apex oxygens with
considerably larger Cu-Cl$_{apex}$ separation  (2.86\AA) than that
of Cu-O$_{apex}$ (2.42\AA) in La$_2$CuO$_4$. Hence, in
Sr$_2$CuO$_2$Cl$_2$ one has the opportunity to examine the CuO$_2$
plane states, both copper and oxygen, without the ''parasitic''
contribution of apex oxygens. At present, there are a rather large
number of experimental data for Sr$_2$CuO$_2$Cl$_2$ obtained with
the help of optical spectroscopy, \cite{optics} X-ray photoemission
(XPS), \cite{x-ray} ultraviolet photoemission (UPS), \cite{Fujimori}
X-ray absorption spectroscopy, (XAS) \cite{XAS,XAS1} and
angle-resolved photoemission spectroscopy.
\cite{LaRosa,Kim,Ronning,Haffner1,Wells}

The 0D cuprates with  isolated CuO$_4$ plaquettes like
Bi$_2$CuO$_4$, the corner-shared and edge-shared 1D chain cuprates
like Sr$_2$CuO$_3$ and Li$_2$CuO$_2$, respectively, as well as the
2D systems like Sr$_2$CuO$_2$Cl$_2$ should reveal similar signatures
of the one-center excitations. However, what concerns the two-center
excitations, there appears a principal difference. Naturally, in
tetragonal 2D system of corner-shared  CuO$_4$ centers, the
two-center excitations manifest itself equally both in $a$ and $b$
in-plane polarizations, while in 1D system of the corner-shared
CuO$_4$ centers they are visible in a similar manner only for
"longitudinal" polarization when the electric field is parallel to
the chain direction. For the electric field parallel to the CuO$_4$
plane but perpendicular to the chain direction one might observe
only $\pi$-like one-center excitations whose intensity is determined
by the overlap contribution only. Thus, by comparing the optical
response for an idealized 1D chain system with two polarizations we
could estimate the contribution of the $\sigma$-like two-center
excitations by simple subtraction of the two spectra. Naturally, it
would be an oversimplification to say that one can deduce the
optical response for an idealized 2D plane by simple combination of
the 1D spectra with different polarization. A correct extension of
the chain data to predict the 2D response should also take into
account the additional $90^o$-coupling of the two-center excitations
already at the $\Gamma$-point and the more complex dispersion
relations. Nevertheless, a simple naive comparative analysis of the
chain and plane optical responses could provide important
semi-quantitative information on the relative contribution of one-
and two-center excitons, their spectral positions, and the relative
magnitudes of the overlap and covalent contributions to the dipole
matrix elements.

One should note the provisional nature of a certain dimensionality
for real cuprates. Even in the so-called 0D systems like
CuB$_2$O$_4$, the CuO$_4$ plaquettes are not really isolated. It
might be surprising that among the best candidates for 0D cuprates
concerning the optical response are such systems like Li$_2$CuO$_2$
or CuGeO$_3$. Usually they are considered as 1D systems with
edge-sharing CuO$_4$ plaquettes according to their crystalline
structure. For instance, at room temperature, CuGeO$_3$ belongs to
the orthorhombic group D$_{2h}^5$ $(Pbmm)$. The unit cell contains
two edge-sharing strongly deformed CuO$_6$ octahedra, with two types
of Cu-O bonds of 2.77 and 1.94 \AA, forming one-dimensional
antiferromagnetic Cu-O chains along the c axis. However, since the
Cu-O-Cu bond angle along the chain is almost $90^o$, the transfer
from one plaquette to its nearest neighbors is strongly suppressed
for the in-plane states. Thus, the in-plane charge excitations are
localized within one plaquette and with respect to the electronic
properties the CuO$_4$ centers in these compounds can be addressed
to be approximately isolated. \cite{Atzkern} This standpoint is
fairly well confirmed in optical and EELS data for Li$_2$CuO$_2$.
\cite{Atzkern,Li}  In what concerns the perpendicular out-of-plane O
$2p_z$ orbitals these systems should be considered as typical 1D
chain systems with O $2p_\pi$ bonding.

In addition, one should note the problem related to the different
symmetry of the crystalline field for the CuO$_4$ plaquettes in 0D
and 2D systems as compared with 1D systems, and between edge-shared
and corner-shared 1D systems. For instance, two inequivalent oxygen
sites are clearly seen in the polarization-dependent O $1s$ x-ray
absorption spectra for the 1D cuprate Sr$_2$CuO$_3$.
\cite{Drechsler} However, the energy separation is found to be as
small as $0.5$ eV, whic is sufficiently smaller than values of the
order $1.5-2.0$ eV, expected from predictions of the simple
point-charge model, or LDA calculations. \cite{Drechsler} This
experimental finding evidences puzzlingly small non-tetragonal
effects near the ground state for the CuO$_4$ centers in the 1D
cuprate Sr$_2$CuO$_3$ and allows to make a comparative analysis of
the localized excitations with those of the tetragonal 2D cuprate
Sr$_2$CuO$_2$Cl$_2$. It seems that the internal Cu-O covalent bond
for a CuO$_4$ plaquette provides the main contribution to its
electronic structure.

\subsection{The main predictions of the model theory}

Before addressing the experimental EELS data we would like to
summarize shortly the main results of the model theory of one- and
two-center CT excitons in 2D insulating cuprates.

\subsubsection{One-center excitons }

i) We predict three types of dipole-allowed one-center excitons: one
exciton $b_{2u}$ with out-of-plane polarization, and two excitons
$e_{u}(\pi)$ and $e_{u}(\sigma)$ with in-plane polarization and
predominantly $\pi$  and $\sigma$ O 2p orbital weight, respectively.

ii) The relative value of transition matrix element  for the
low-energy $e_{u}(\pi)$ exciton with an estimated energy $\approx 2$
eV and for the high-energy $e_{u}(\sigma)$ exciton is determined by
the level of the $\pi -\sigma$ mixing.

iii) The low-energy (near 2 eV) part of the absorption spectrum for
a single CuO$_4$ plaquette is formed by an interplay of forbidden
$d-d$ transitions ($b_{1g}^b \rightarrow b_{2g}^b, a_{1g}^b,
e_{g}^b$), forbidden  $b_{1g}^b \rightarrow a_{2g}$ and allowed
$b_{1g}^b \rightarrow e_{u}(\pi)$ CT transitions, respectively,
which are all close in energy. It is worth noting that for cuprates,
as for many other strongly covalent oxides, the conventional
division of optical transitions into crystal-field $d-d$ and CT
$p-d$ transitions becomes questionable, since the $b_{1g}^b
\rightarrow b_{2g}^b, a_{1g}^b, e_{g}^b$ transitions are accompanied
by a strong $p-d$ and $p-p$ charge transfer. The low-energy
absorption band is expected to be a result of strong
electron-vibrational coupling which lifts the selection rules for
certain transitions.

iv) The final $e_{u}(\pi)$ and $e_{u}(\sigma)$ states are unstable
with respect to the formation of Jahn-Teller (or pseudo-Jahn-Teller)
centers with localization of the appropriate  CT excitation. This
effect is especially important for the $b_{1g}^{b} \to e_{u}(\pi)$
excitation, because the predominant $\pi$ nature of the hole state
results in small transfer integrals and a relatively large effective
mass for the exciton. The localization is accompanied by the
vibronic reduction of transition matrix elements. The high-energy
$e_{u}(\sigma)$ exciton is expected to manifest significant
intensity both in optical and EELS spectra.

\subsubsection{Two-center excitons }

i) All the two-center  excitons generated by the CT governed by the
strongest $\sigma$ bonds can be divided into three channels
$b_{1g}^2, b_{1g}a_{1g},b_{1g}e_{u}$, respectively, in accordance
with the symmetry of the final two-hole state. The main effects in
optics and EELS are associated with the $b_{1g}^{2}$- and
$b_{1g}e_{u}$-channels. In the frame of each channel we deal with
different final two-hole states whose energy and wave functions are
mainly determined by correlations, or configuration interaction
effects. The different theoretical model calculations predict the
lowest energy of two-center CT transition for the
$b_{1g}^{2}$-channel to be of the order $2\div 3$ eV.

ii) All the two-center  excitons could be divided to even $S$- and
odd $P$-excitons. For 2D cuprates these excitons generate two sets
of four $A_{1g},B_{1g},E_{u}$  plane modes ($S,D,P_{x,y}$ modes).
\cite{Ng}

iii) The transition matrix elements in optics and EELS have rather
complicated ${\bf k}$ dependence, substantially differing for the
oxygen and copper contributions. The simple dipole approximation
fails to explain correctly the ${\bf k}$-dependent effects in EELS.

iv) In frame of the $b_{1g}^{2}$-channel we predict two most
important excitons with an energy separation $\approx 7.0 $ eV. The
low-energy exciton with ZR-singlet as a final hole state has rather
unusual ${\bf k}$ dependence of intensity along the $(\pi , \pi)$
direction with  a sharp decrease, or even compensation point, by
going to the BZ boundary. The high-energy exciton with final
ZR-singlet-like ${}^{1}A_{1g}$ state of predominantly O 2$p$ nature
is likely to manifest the largest intensity for the
$b_{1g}^{2}$-channel.

v) In frame of the $b_{1g}e_{u}$-channel we predict four most
important excitons, generated by CT to $e_{u}(\pi)$ and
$e_{u}(\sigma)$ orbitals with smaller and larger weight of O
2$p_\sigma$ orbitals, respectively. These weights determine the
relative intensities of the appropriate excitons. The photoemission
data \cite{Pothuizen,Duerr} predict the energy separation between
the low-lying $b_{1g}^{b} \to e_{u}(\pi)$ and $b_{1g}^{b} \to
b_{1g}$ transitions to be of the order $1.5\div 2.0$ eV. The
$b_{1g}e_{u}$ excitons manifest a rather simple dispersion of
intensity along both main directions in BZ zone with gradual
interchange between dipole allowed and forbidden modes.

\subsubsection{CT excitons and magnetic subsystem}

All parent 1D and 2D copper oxides with corner-shared CuO$_4$
plaquettes are  antiferromagnets with spin-$1/2$, localized on Cu
atoms of the chains or planes, with exchange integral $I\approx 0.1$
eV and a N\'{e}el temperature $T_N \sim 300$ K in 2D. So, we have to
address the excitons in antiferromagnets. Historically, this problem
has been intensively studied in connection with forbidden $d-d$
excitonic transitions in weakly covalent antiferromagnetic
insulators like MnF$_2$ (see e.g. the paper by Tanabe and Aoyagi in
Ref.\ \onlinecite{Sturge}). At first glance, the main allowed
electric dipole  transitions preserve the spin state and cannot
result in conventional or orientational spin fluctuations. On the
other hand, the charge transfer transitions are accompanied by a
strong spatial redistribution of spin density, and, in a sense,
result in a spin density fluctuations. Indeed, the two-center CT
exciton is formally associated with spin singlet-singlet transition
from a spin-singlet initial state with a $nonzero$ spin density on
both centers to a spin-singlet final state with a $zero$ spin
density on both centers. In a sense, this transition is accompanied
by a local annihilation of the N\'{e}el-like or spin-dimerized
state, and the creation of a non-magnetic vacancy. Such a strong
spin density fluctuation could result in effective two-magnon (2M)
processes. In other words, one might expect strong 2M Raman
scattering generated by a two-center exciton. Indeed, the
experimental Raman measurements for different insulating cuprates
\cite{Blumberg,Heyen} show that the maximal strength of 2M
scattering occurs for excitation energies substantially higher than
the optical band edge with a peak which we assign to a two-center
exciton.

For small one-center excitons we have a rather conventional $s=1/2
\rightarrow s=1/2$ transition with spin density fluctuation
localized inside the CuO$_4$ plaquette. Such a transition is not
accompanied by strong 2M Raman processes, what could be used to
identify such kind of excitons. The redistribution of spin density
from the copper atom to the oxygen ones for the $b_{1g}\rightarrow
e_u$ transition switches on the strong ferromagnetic Heisenberg O
$2p$-Cu $3d$ exchange with the nearest neighbor CuO$_{4}$
plaquettes. Interestingly, the different sign of exchange coupling
for $e_u$ and $b_{1g}$ holes with the same neighborhood,
ferromagnetic  for the former, and antiferromagnetic for the latter,
leads to a number of temperature anomalies near the 2D-3D
antiferromagnetic phase transition. There is at first the $blue$
shift effect for the energy of the small one-center exciton $b_{1g}
\to e_u$ by lowering the temperature near and below $T_N$. Indeed,
at $T \gg T_N$, the average molecular field for the CuO$_4$ center
turns to zero. The 3D antiferromagnetic ordering is accompanied by a
rise of exchange molecular fields and respective spin splittings.
Due to the different signs  of molecular fields for $e_u$ and
$b_{1g}$ states this is accompanied by an increase of the transition
energy with maximal value of $blue$ shift $\Delta \approx \beta
(|H_{b_{1g}}|+|H_{e_{u}}|)$. This quantity could be as large as
several tenths of eV. Additionally, one has to expect a strong (of
the same order of magnitude) broadening of the excitonic line with
increase of the temperature due to strong fluctuations of molecular
fields. All these expectations are experimentally found for the
$2.0$ eV line in Sr$_2$CuO$_2$Cl$_2$ \cite{Choi} confirming its
one-center excitonic nature. A similar situation is observed in
La$_2$CuO$_4$,\cite{Falck} although the authors have explained the
data by assuming a polaronic nature of electrons and holes with a
short-range interaction in between.

It should be noted that at present there is no relevant theory of
spin-excitonic coupling for small CT excitons in strongly covalent
and correlated systems. The traditional framework for understanding
the two-magnon Raman scattering in  antiferromagnets for
non-resonant excitation energies has been an effective Loudon-Fleury
Hamiltonian,\cite{Loudon} which implies  well localized spins and
weak spin-wave like spin fluctuations.

\subsubsection{CT excitons and phonon subsystem}

Generally speaking, the CT exciton creation is usually accompanied
by an excitation of lattice modes. Indeed, the hole transfer from Cu
$3d$ to O $2p$ state, or from an $ionic$ to a $covalent$
configuration is accompanied by a significant shortening of the
equilibrium Cu-O bond length. The exciton-phonon interaction
strongly influences the line-shape of absorption and results in a
phonon Raman scattering. The measurement of the Raman intensity as a
function of excitation light energy is a very informative probe of
the origin of electronic transitions.\cite{Heyen} In particular,
this method has allowed \cite{Ohana} to resolve a fine structure of
the low-energy (LE) excitonic feature in La$_2$CuO$_4$ with a sharp
peak at $2.14$ eV as narrow as 50 meV and a broader structure at
$1.9$ eV. In addition, the authors made the important observation of
the selective resonance enhancement, namely that neither the
first-order even-parity phonon nor their harmonics show resonant
enhancement near the LE exciton peak, while the high-energy (above
450 cm$^{-1}$) presumably odd Raman modes show large resonant
enhancement of as much as a factor of $10-40$ relative to a laser
energy of $2.7$ eV. Similar results have been obtained for
insulating YBa$_2$Cu$_3$O$_{6.1}$.\cite{Heyen} In our opinion, such
an unusual behavior of the LE exciton could be associated with its
Jahn-Teller nature. Indeed, the $b_{1g}\rightarrow e_u$ transition
to the orbital doublet $e_u$ represents a textbook example of a
so-called $A-E$ transition.\cite{bersuker} The excited, orbitally
degenerate $e_u$ state is unstable  with regard to vibronic coupling
with the local distortion modes $A_{1g},B_{1g},B_{2g}$, $E_u$ and
the formation of a polaron-like (soliton-like) vibronic center with
a complex two- or four-well adiabatic potential 
and a rather strong renormalization of vibration frequencies. In
other words, the doublet $e_u$ state tends to a spontaneous local
symmetry breaking, including removal of the inversion center due to
interaction with the close in energy even states and
pseudo-Jahn-Teller effect. Namely the latter would result in a
strong resonance coupling of the  $b_{1g} \to e_u$ exciton with odd
$E_u$ lattice modes. Naturally, the structure of such a center would
strongly depend on differences in the bare lattice and elastic
parameters and differ in 214 and 123 systems. One should emphasize
that the formation of the heavy polaron-like, or $localized$ small
exciton would result in a strong enhancement of its effective mass.

\subsection{Cuprates with well isolated or weakly coupled CuO$_4$
plaquettes (0D cuprates)}

\begin{figure}[h]
\includegraphics[width=8.5cm,angle=0]{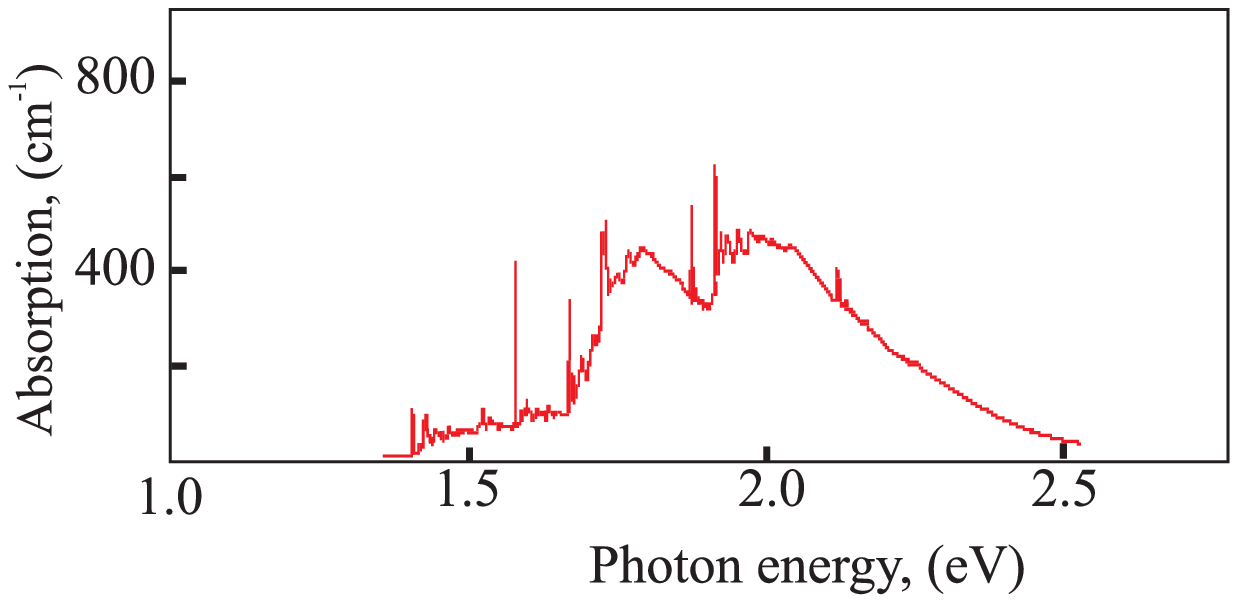}
\caption{Low-energy absorption spectra for the 0D cuprate
CuB$_2$O$_4$($\bf k \parallel z, \bf E \perp z$).\cite{Pisarev}} \label{fig6}
\end{figure}

In Fig.\ \ref{fig6} we reproduce the low-energy absorption spectra for
the 0D cuprate CuB$_2$O$_4$.\cite{Pisarev} In contrast to most 1D
and 2D cuprates where  the low-energy absorption bands are broad and
featureless, the absorption spectra for the true 0D cuprate
CuB$_2$O$_4$ (Fig.\ \ref{fig6}) provide the remarkable opportunity to
observe the interplay of low-energy allowed and forbidden
transitions.  The experimental spectrum shows a number of narrow
zero-phonon peaks with up to 70 well-resolved phonon
sidebands\cite{Pisarev} imposed on  rather broad bands. The former
may be assigned to forbidden $b_{1g}^b \rightarrow b_{2g}^b,
a_{1g}^b, e_{g}^b$ ($d-d$) and $b_{1g}^b \rightarrow a_{2g}$ ($p-d$)
transitions, while the latter may be assigned to allowed $b_{1g}^b
\rightarrow e_{u}(\pi)$ CT transitions in two types of CuO$_4$
plaquettes, respectively. The transition energies at Cu$_{4b}$O$_4$
and Cu$_{8d}$O$_4$ plaquettes are similar because of the small
influence of the remote apical O$^{2-}$ ions. The experimental EELS
spectra for CuB$_2$O$_4$ \cite{Knupfer} exhibit a rather broad
dispersionless band peaked near $5$ eV  which could be associated
with a phonon-assisted {\it localized electron excitation}
$b_{1g}^{b}\rightarrow e_{u}^{a}$. The decrease of intensity with
the increase of momentum agrees with the cluster (CuO$_4$) model for
electronic states  and the dipole nature of transition. A weak
feature distinctly observed at small momenta near $2\div 3$ eV could
be assigned to the {\it localized electron excitation}
$b_{1g}^{b}\rightarrow e_{u}^{b}$, i.e.\ to the low-energy bonding
counterpart of $e_{u}$ state.
The optical and EELS spectra for 0D cuprates are governed only by
the intra-center transitions.

Interestingly, that very similar EELS spectra are observed in
Sr$_2$CuO$_3$ for polarization ${\bf q}\parallel {\bf b}$, that is
perpendicular to the chain direction ({\it out-off-chain} spectra,
\cite{Neudert} Fig.\ \ref{fig7}). Indeed, like for 0D systems, there
are no two-center $eh$-excitations governed by strong covalent
$\sigma$-bonding in this case, and we deal with the predominant
contribution of intra-center excitations. So, for the polarization
perpendicular to chain direction, the 1D cuprates like Sr$_2$CuO$_3$
provide the optical and EELS response typical for a 0D system.
Contrary to CuBi$_2$O$_4$, both dipole-allowed
$b_{1g}^{b}\rightarrow e_{u}^{b,a}$ transitions in Sr$_2$CuO$_3$
manifest itself more distinctly with well-defined EELS-peaks at 2.0
and 5.4 eV, respectively. The appropriate peaks in optical
conductivity are situated at 1.8 and 4.3 eV, respectively. One
should note that these EELS data present a straightforward
experimental manifestation of the dipole-allowed intra-center
$b_{1g}^{b}\rightarrow e_{u}^{b,a}$ excitations without the
"parasitic" effect of inter-center transitions.

The situation in 1D systems crucially changes for polarization
parallel to the chain direction, or for the {\it in-chain} spectra.
Even in Li$_2$CuO$_2$, where, similarly to real 0D systems, we deal
with a predominant contribution of intra-center transitions, the
{\it in-chain} EELS spectra \cite{Atzkern} look substantially
different from those for CuBi$_2$O$_4$. First of all, it concerns
the line-shape of the intensive band in the spectral range $4\div 5$
eV with a very sharp peak at 4.7 eV ($\Gamma$-point) and  small
negative dispersion, which evidences its excitonic nature. Indeed,
the concept of the one-center $b_{1g}^{b} \to e_{u}^{a}$ exciton was
successfully applied \cite{Atzkern} for a quantitative description
of these EELS spectra.

Similar absorption spectra as shown in Fig.\ \ref{fig6} are also
observed in 1D cuprates with 90$^{\circ}$ Cu-O-Cu bonds like
CuGeO$_3$. \cite{Bassi}
The EELS spectra of Li$_2$CuO$_2$ present another example for the
weak energy dispersion  of the one-center exciton accompanied by a
strong dispersion of its intensity. The different behavior of the
one-center dipole-active $b_{1g}^{b}\rightarrow e_{u}$ excitations
with {\it in-chain} and {\it out-off-chain} polarization is
straightforwardly associated with specific translational symmetry
and lattice dynamics of 1D systems. Indeed, only {\it in-chain}
component of momentum is conserved and can describe the electron and
lattice modes. Additionally, the 1D systems are soft with respect to
{\it out-off-chain} distortions that favor the localization of
electron excitations. The most probable mechanism governing the
structure of the polaron-like one-center exciton  in Li$_2$CuO$_2$
is associated with strong coupling of the purely oxygen O
2$p_\sigma$ hole mode and oxygen distortion modes at the BZ
boundary.

\subsection{Polarization dependent EELS spectroscopy of 1D copper oxide
Sr$_2$CuO$_3$ and separation of one- and two-center CT excitons.}

In 2D systems we usually deal with spectra being a hardly resolved
superposition of both types of excitons. This renders the separation
of transitions with different nature rather difficult. Therefore,
there remains still some ambiguity concerning the reliable
identification of $two$ dipole-allowed one-center CT excitons and
$seven$ essential two-center CT excitons  that might still question
its existence as well-defined entities. Unfortunately, this concerns
also the structure of the low-energy optical response observed in
the spectral range $2\div 3$ eV, which is of special importance
since it is associated with the states which are believed to define
the unconventional properties of the cuprates. One should emphasize
that conventional optical measurements are restricted only to the
$\Gamma$-point and cannot separate between localized one-center,
dispersionless excitations and two-center excitons with a noticeable
dispersion. In contrast to optics, the angle-resolved EELS
spectroscopy provides unique opportunities to reveal the exciton
dispersion and separate one- and two-center CT excitons.
One-dimensional cuprate compounds are good candidates for such a
study since the scattering of electrons with a transferred momentum
${\bf q}$ perpendicular to the chain direction excites only
electron-hole pairs sitting on one CuO$_4$ plaquette. On the other
hand, for ${\bf q}$ parallel to the chain direction, both types of
excitons (one- and two-center) can be observed. The interpretation
of the EELS spectra with ``longitudinal'' response were already
proposed in terms of the standard one-band Hubbard model.
\cite{1D-EELS}
Below we show that the polarization-dependent angle-resolved EELS
study of the 1D cuprate Sr$_2$CuO$_3$ with corner-shared CuO$_4$
plaquettes  provides a unique opportunity  to separate both the
one- and two-center CT excitons, and the two types of one-center
excitons, with the first unambiguous manifestation  for the
relevance of  the O $2p_{\pi}$ holes for the low-energy excitations
in cuprates.

The EELS spectra for Sr$_2$CuO$_3$ in "longitudinal" response with
the transferred momentum oriented along the chain direction were
measured earlier on \cite{1D-EELS} and have been interpreted within
standard Hubbard models. However, such models can describe properly
only the "longitudinal" response with the transferred momentum
oriented along the chain direction. In this paper we argue that
along with the "longitudinal" response a 1D cuprate compound like
Sr$_2$CuO$_3$ should reveal rather strong low-energy "transversal"
response for the transferred momentum oriented in the plane of the
CuO$_4$ plaquettes but perpendicular to the chain direction.
\begin{figure}[h]
\includegraphics[width=8.5cm,angle=0]{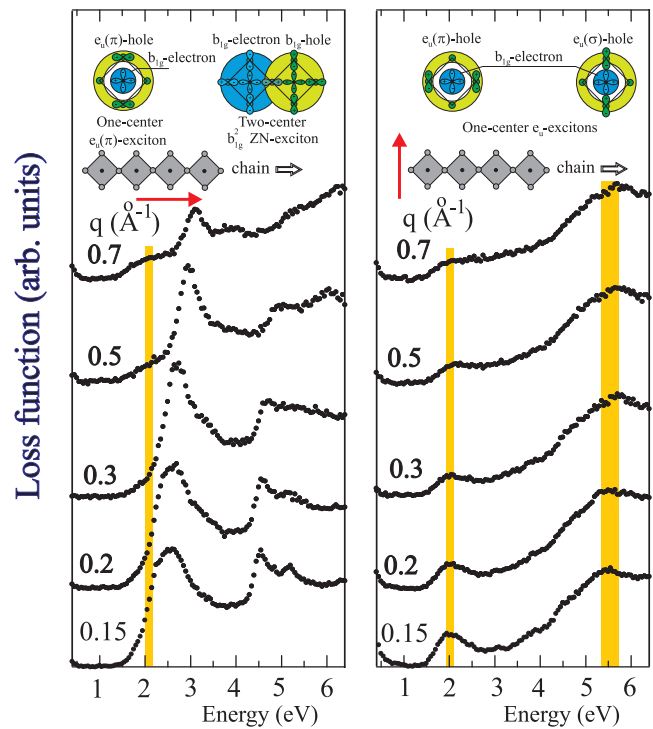}
\caption{EELS spectra  in Sr$_2$CuO$_3$ for longitudinal ${\bf
q}\parallel {\bf a}$ (left panel)
 and transversal ${\bf q}\parallel {\bf b}$ (right panel)
  response with an illustration of  one-
and two-center excitons. } \label{fig7}
\end{figure}
The EELS spectra for Sr$_2$CuO$_3$ in both polarizations are
presented in Fig. \ref{fig7}. Their comparison leads to very
important conclusions. As is to be expected, we see a strong
difference between the two sets of spectra. First, this concerns the
well-defined dispersionless EELS peaks at 2.0 and 5.5 eV in
transversal polarization (right hand side panel in Fig.\
\ref{fig7}). The intensity considerations and the absence of
noticeable energy dispersion allow us to associate them  with
dipole-allowed CT transitions having a particularly localized
nature. Moreover, when comparing both spectra we see that despite
the strong inequivalence of longitudinal and transversal
polarizations in Sr$_2$CuO$_3$ the low-energy  transition peaked
near 2 eV is equally present in both polarizations. Though for
longitudinal polarization this is partly hidden for low momentum
values by the intensive band peaked at 2.6 eV and is seen as a
shoulder,  near the BZ boundary it is a well-separated weak band due
to the big blue-shift of the intense neighbor. This fact implies
that the excitation is localized on the single CuO$_4$ plaquette
being the only common element of longitudinal and transversal
geometry in this 1D cuprate with corner-shared CuO$_4$ plaquettes.
Hence, by taking into account the intensity ratio we can
unambiguously  identify the 2.0  and 5.5 eV peaks in the EELS
spectrum of Sr$_2$CuO$_3$ with the one-center CT excitons
$e_{u}(\pi)$ and $e_{u}(\sigma)$, respectively.

The sizeable dispersion  of the most intense low-lying CT exciton
peaked in EELS at 2.6 eV agrees with its two-center nature
\cite{Wang,Ng} that is  fairly well confirmed in the studies of
nonlinear optical effects in Sr$_2$CuO$_3$. \cite{Ogasawara,Kishida}
Indeed, our analysis of the photoinduced absorption experiments and
the measurements of the third-order nonlinear susceptibility $\chi
^{(3)}$ in Sr$_2$CuO$_3$ \cite{Ogasawara,Kishida} allows to reveal
the near-degeneracy of both $S$- and $P$-types of two-center CT
excitons with an energy of about 2.0 eV, and  to obtain a reliable
estimate for the transition matrix element (\ref{me}): $|\langle
S|\hat{\bf d}|P\rangle |\approx 2e\times 4$ \AA \ (see Fig.\ 4 in
Ref.\ \onlinecite{Kishida}), that straightforwardly points to a
two-center CT exciton. In addition, the nonlinear optical response
of Sr$_2$CuO$_3$  \cite{Kishida} reveals a weak spectral feature
red-shifted to several tenths of eV with regard to the position of
the main two-center exciton. It can be attributed to a weak
one-center exciton that seems yet to be $invisible$ in reflectivity
measurements of Sr$_2$CuO$_3$. \cite{Tokura} In this connection, we
would like to emphasize once more the decisive role of $direct$ EELS
measurements in the observation and unambiguous assignment of one-
and two-center excitons in Sr$_2$CuO$_3$ as compared with
conventional $indirect$ optical data.

It is interesting to note that a low-energy transition with small
dispersion was also seen in the RIXS data of the 1D cuprate
SrCuO$_2$ published in Ref.\ \onlinecite{Kim04}. In that paper it
was interpreted as the onset of the continuum of the one-band
Hubbard model. By comparing it with the above discussion of the EELS
data of Sr$_2$CuO$_3$ we are led to an alternative interpretation of
these RIXS data in terms of a superposition of two different
electron-hole excitations, namely the above discussed one- and
two-center excitons.

\subsection{CT excitons in  2D insulating cuprates}

At first sight, all the above findings are restricted to 1D
cuprates. However, already a shorthand inspection of the
high-resolution EELS spectra  for the 2D system Sr$_2$CuO$_2$Cl$_2$
for two polarizations \cite{EELS,EELS1} shows nearly the same
behavior of the lowest in energy electron-hole excitations. Being
encouraged by such an impressive success of our model we can now
address the detailed analysis of CT excitons in  2D insulating
cuprates.

\subsubsection{EELS  spectra of Sr$_{2}$CuO$_{2}$Cl$_{2}$}

As it was already noted, Sr$_2$CuO$_2$Cl$_2$ is one of the best
realizations of a 2D antiferromagnetic, insulating model compound.
Of particular importance for the assignment of both one- and
two-center excitons are data obtained by ARPES which is a powerful
tool to examine the energy spectrum of the two-hole CuO$_{4}^{5-}$
center. It probes one-particle excitations in contrast to optics and
EELS, and the nonbonding oxygen states are clearly visible. At the
$\Gamma$-point the selection rules for ARPES are the same as for
optical and EELS transition, so that ARPES "sees" only the purely
oxygen $e_u$ photoholes, or the ${}^{1}E_u$ states of the two-hole
CuO$_{4}^{5-}$ center with $b_{1g}e_u$-like configuration. The
inspection of the experimental ARPES spectra for
Sr$_{2}$CuO$_{2}$Cl$_{2}$ obtained in Refs.\
\onlinecite{Pothuizen,Duerr} gives valuable information regarding
the two low-lying ${}^{1}E_u$ states. It clearly reveals two strong
bands separated from the ground state ZR-singlet by $1.5\div 2.0$ eV
and $\approx 5.0$ eV, which could be naturally related to the
low-energy $e_{u}(\pi)$ and high-energy $e_{u}(\sigma)$ photohole
states, respectively. Different theoretical estimations
\cite{Hayn,McMahan1,Mattheiss,Tanaka,Tanaka1} corroborate the ARPES
data and point to a rather low-energy position of the O $2p_{\pi}$
states in 1D and 2D cuprates with a corner shared arrangement of
CuO$_4$ plaquettes with $\Delta_{\sigma \pi}=(\epsilon_{p
\sigma}-\epsilon_{p \pi})\approx 1\div 3$ eV, where $\epsilon_{p
\sigma}$ and $\epsilon_{p \pi}$ are the centers of gravity for
different O $2p_{\sigma}$ and O $2p_{\pi}$ states, respectively.

Unfortunately, the measurements in Ref.\ \onlinecite{Duerr} are
energy restricted and do not provide information regarding the
high-energy $b_{1g}^{a}e_{u}(\pi)$ and $b_{1g}^{a}e_{u}(\sigma)$
configurations. However, recent resonant X-ray scattering
spectroscopy (RIXS) measurements for different insulating cuprates
\cite{Abbamonte,Hasan,Hill} with the incident photon energy tuned to
the Cu $K$ edge (hard X-rays) reveal a rather wide ($2\div 3$ eV)
band peaked at $5\div 6$ eV, which could be unambiguously identified
as the band of antibonding Cu $3d$-O $2p$
($a_{1g}^{a},b_{1g}^{a},b_{2g}^{a},e_{g}^{a}$) states. Hence, the
$b_{1g}^{b}-b_{1g}^{a}$ separation could be estimated to be $\approx
6.0$ eV.

Thus, making use of the ARPES and RIXS data, as well as the
theoretical predictions, we can establish an overall picture of
optical and EELS spectra (at the $\Gamma$-point). These spectra are
governed by dipole-allowed one- and two-center CT excitations and
they include:

i) the lowest in energy ($1.5\div 2.0$ eV) one-center $b_{1g}^{b}
\to e_{u}(\pi)$ transition with rather small intensity due to the
dominantly O $2p_\pi$ nature of the final state and strong tendency
to self-localization (trapping);

ii) the high-energy ($\approx 5.0$ eV) and rather intensive
one-center $b_{1g}^{b} \to e_{u}(\sigma)$ transition;

iii) in frame of the $b_{1g}^{2}$-channel we predict three
two-center CT transitions: 1) the rather intensive lowest in energy
($2.0\div 3.0$ eV) transition with ZR-singlet as a final hole state;
2) the most intensive high-energy transition with ZR-singlet-like
final hole state blue-shifted to $\approx 7.0$ eV; and 3) probably
the less intensive transition blue-shifted to $\approx 5.0$ eV with
regard to the first one;

iv) in frame of the $b_{1g}e_{u}$-channel we predict two doublets of
two-center CT transitions shifted $\approx 6.0$ eV with regard to
each other. The high-energy doublet is relatively more intensive.
The partial transitions in both doublets are shifted $3.0\div 3.5$
eV with regard to each other. Among them, the lowest transition
which is rather intense to the $b_{1g}e_{u};dp \pi$ state is
blue-shifted by $1.5\div 2.0$ eV with regard to the lowest in energy
$b_{1g}^{2}$ transition with ZR-singlet as a final hole state, while
the more intensive high-energy $b_{1g}e_{u};dp \sigma$ transition is
blue-shifted by $\approx 5.0$ eV with regard to the same transition.
In this quartet the most intensive CT transition with
$b_{1g}e_{u};pp \sigma$ final state is expected to have  the maximal
energy ($\sim 13\div 14$) eV among all the CT transitions governed
by the strong $\sigma$ bond.

\begin{figure}[h]
\includegraphics[width=8.5cm,angle=0]{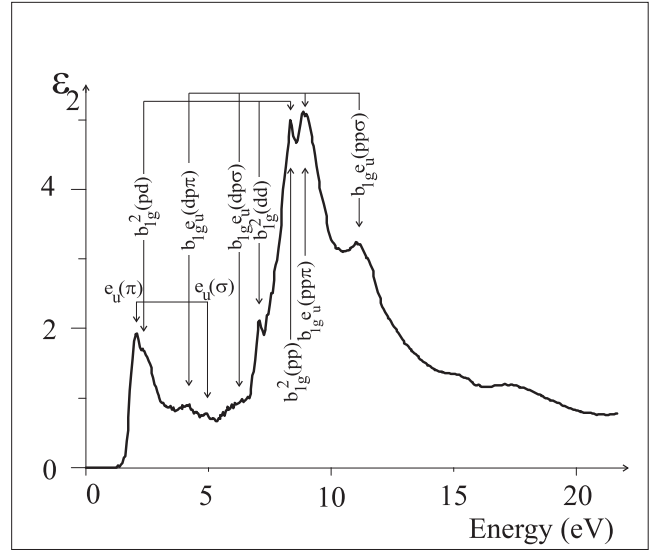}
\caption{The spectral dependence of the imaginary part of
dielectric function $\epsilon (\omega)$ for Sr$_2$CuO$_2$Cl$_2$.
Arrows mark
 the predicted energy position of one-center, two-center
 $b_{1g}^2$-channel, and $b_{1g}e_u$-channel
 CT transitions.} \label{fig8}
\end{figure}

As we shall see, all these predictions are  in  agreement with
experimental optical and EELS spectra available for
Sr$_2$CuO$_2$Cl$_2$. \cite{Neudert}

In Fig. \ref{fig8} we present the spectrum of the  imaginary part of
dielectric permittivity derived from the Kramers-Kronig
transformation of the EELS data.\cite{Neudert} Remarkably, all the
excitations which were analysed before correspond to visible
features in $\epsilon_2$. The weak low-energy feature near $2.2$ eV
in the EELS spectra ($2.0$ eV in $\epsilon_{2}$ and optical
conductivity) could be ascribed to the lowest in energy  one-center
$b_{1g}^{b} \to e_{u}(\pi)$ transition. This feature occupies the
tail of a rather intensive band peaked at $2.7$ eV in EELS spectra
($2.4$ eV in $\epsilon_{2}$ and optical conductivity). This band is
naturally associated with the  lowest in energy $b_{1g}^{b} \to
b_{1g}$ two-center CT transition with ZR-singlet as a final hole
state. Its energy is of particular importance for the whole set of
two-center transitions. Having positioned the lowest one- and
two-center excitons, the higher states are fixed by the chosen
parameter values (see Fig.\ \ref{fig2}). Two features in EELS, near
$4.2$ eV and $7.1$ eV could be related to two-center $b_{1g}e_{u};
dp \pi$ and $b_{1g}e_{u}; dp \sigma$ transitions, respectively,
while the wide band near $6.0$ eV in EELS could be naturally
assigned to the one-center $b_{1g}^{b} \to e_{u}(\sigma)$
transition, whose energy is slightly higher than that of its
two-hole counterpart. The overall spectrum of the most effective CT
transitions generated by the Cu $3d$-O $2p$ $\sigma$ transfer ends
with very strong features  near $9\div 10$ and $12\div 13$ eV in
EELS which could be certainly assigned to the high-energy two-center
$b_{1g}^{2}; pp$  CT transition  with ZR-singlet-like  final hole
state of predominantly $pp$ configuration, and the transitions to
the $b_{1g}e_{u}; pp \pi$ and $b_{1g}e_{u}; pp \sigma$ final states,
respectively. All this shows that important spectral information is
contained in the range above 8 eV. The peak at about 18 eV is likely
to be attributed to transitions with O $2s$ initial state.

\begin{figure}[h]
\includegraphics[width=8.5cm,angle=0]{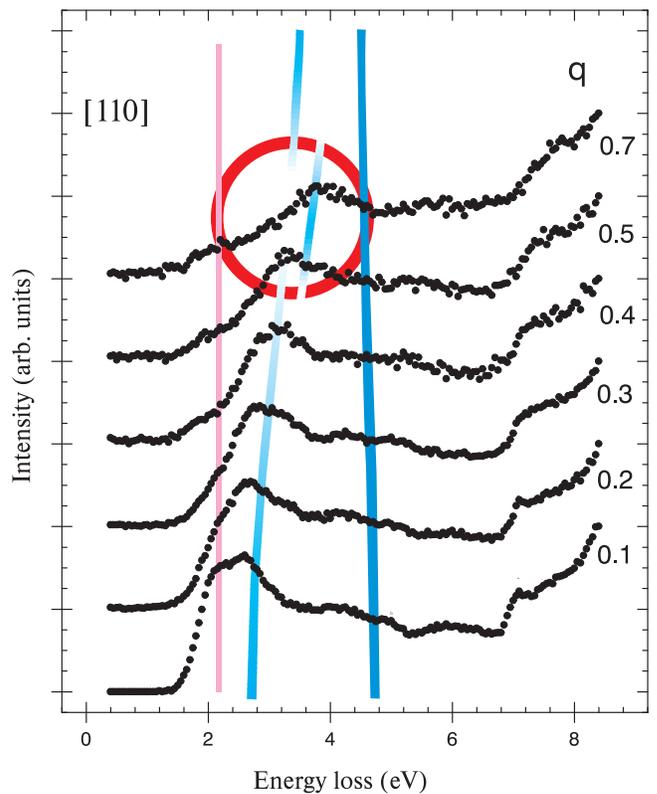}
\caption{The spectral dependence of the loss function for
Sr$_2$CuO$_2$Cl$_2$. The region of possible destructive
exciton-exciton interference is highlighted by a circle. See text
for details.} \label{fig9}
\end{figure}

The interpretation of EELS spectra for nonzero momentum becomes more
complicated due to the energy and intensity dispersion of the
dipole-allowed modes and the appearance of numerous new modes which
are forbidden at the $\Gamma$-point. Hereafter, we focus only on
angle-resolved EELS spectra in the spectral range up to $8.0-8.5$ eV
for several momentum values in $[100]$ and $[110]$ directions,
obtained by Wang {\it et al.}\cite{Wang} and by Fink {\it et
al.}\cite{EELS,EELS1} The spectra differ somewhat by the maximal
momenta values and the considerably better resolution in the latter
case. For illustration, we reproduce in Fig.\ \ref{fig9} the EELS
spectra \cite{EELS,EELS1} along the $[110]$ direction.
The low-energy part of the EELS spectra in the 2D system
Sr$_2$CuO$_2$Cl$_2$ along the $[110]$ direction and the longitudinal
spectra in the 1D system Sr$_2$CuO$_3$ have a very similar structure
up to some quantitative coincidence. For both compounds the main
spectral feature is associated with an intensive band related to the
two-center $b_{1g}^{2}$ ZN-exciton with clear dispersion peaked at
2.6 eV near the $\Gamma$-point and at 3.0 eV near the BZ boundary.
In both systems this band has the low-energy shoulder near $2.0$ eV
which is distinctly seen in the $\Gamma$-point or near the BZ
boundary. This two-peak structure of the CT band in
Sr$_2$CuO$_2$Cl$_2$ is corroborated by conventional optical
measurements \cite{optics,Choi} (see Fig.\ \ref{fig10}).
The EELS spectra along the $[100]$ direction manifest an expected
overall drop in intensity with relatively small energy dispersion.
In contrast to the $[100]$ direction, the main peak for the $[110]$
direction of Sr$_2$CuO$_2$Cl$_2$ exhibits clear signatures of strong
dispersion, particularly for momentum values in the range $(0.5\div
0.7)k_{max}$. Namely this effect was a starting point for the model
theory of the CT excitons by Zhang-Ng.\cite{Wang,Ng} However, in our
opinion, these signatures are not only related to the strong energy
dispersion of the ZN-exciton, but also to the unconventional
behavior of the intensity dispersion for different excitons with the
pecularities especially in the momentum range $(0.5 \div
1.0)k_{max}$ shown in Fig.\ \ref{fig5}. Indeed, as it was emphasized
above, the intensity of the main peak in the low-energy part of the
EELS spectrum for Sr$_2$CuO$_2$Cl$_2$, assigned to the
dipole-allowed $P$-exciton associated with the $b_{1g}^{b} \to
b_{1g}$ two-center CT transition with ZR-singlet as a final hole
state, sharply decreases with the increase of momentum along $[110]$
direction with a probable compensation point near the BZ boundary.
This circumstance allows to clearly observe a sharp rise of the
spectral weight in a rather broad range with the well-defined peak
in EELS near 3.8 eV. This spectral feature may be unambiguously
associated with the dipole-forbidden $A_{g}$ counterpart of
two-center $b_{1g}e_{u};dp \sigma$ exciton, which is allowed and
rather intensive at the $(\pi ,\pi )$ point. Interestingly that the
result of this specific behavior of the EELS intensity for different
excitons might be mistaken for manifestation of the energy
dispersion of the main peak associated with the $g-u$ dipole-allowed
two-center $b_{1g}^{2}$ exciton, as it was made by Wang {\it et
al}.\cite{Wang} Naturally, this mistake leads to a conclusion about
very large ($\sim 1.5$ eV) energy dispersion for this exciton. The
real energy dispersion for different excitons in our opinion may not
exceed the reasonable values of the order of $0.5\div 1.0$ eV.

Thus, the overall analysis of the experimental EELS spectra in  the
model 2D insulating cuprate Sr$_2$CuO$_2$Cl$_2$ allows us to
certainly assign a set of distinctly observed features in a rather
wide spectral range $2.0\div 14.0$ eV to the predicted one- and
two-center CT excitons, and confirm the validity of the theoretical
concept based on an embedded CuO$_4$ cluster model.

Interestingly, that the experimental data available allow us to
estimate the difference in the $el$-$h$-bonding energies $E_b$ for
the low-energy one- and two-center CT excitons. Indeed,
$$
E_{b}(e_{u}(\pi
))-E_{b}(b_{1g}b_{1g})=E({}^{1}E_{u})-E({}^{1}A_{1g})
$$
$$
-(E(e_{u}(\pi ))-E(b_{1g}b_{1g}))\approx (2\pm 0.5) \, \mbox{eV},
$$
where the first term in the right hand side represents the energy of
$b_{1g} \to e_{u}(\pi )$ excitation in the hole CuO$_{4}^{5-}$
center with ZR ground state, while the second one represents the
difference in the energies of the respective excitons. For the
estimate we have made use both of ARPES and EELS data.

\subsubsection{Fundamental absorption band in parent cuprates: Interplay of
one- and two-center CT excitons}

A close examination of other materials shows that the two-component
structure of the CT gap appears to be  a common place for  all
parent cuprates (see $e.g. $, Refs.\ \onlinecite{Cooper}, \onlinecite{Falck}, and \onlinecite{Krich}).
The dipole-allowed localized $b_{1g}\rightarrow e_{u}^b$ excitation
within the CuO$_4$ plaquette related essentially to the $e_{u}(\pi)$
state is distinctly seen in optical and EELS spectra for different
insulating cuprates as a separate weak feature or a low-energy
shoulder of the more intensive band near $2.5$ eV assigned to the
inter-center CT transition associated with the Zhang-Rice
singlet-like excitation $b_{1g}^{2};pd$.

Both components are characterized by a different coupling to the
magnetic and phonon subsystems thus providing additional ways to
separate them. For small one-center excitons we have a rather
conventional $s=1/2 \rightarrow s=1/2$ transition with a spin
density fluctuation localized inside the CuO$_4$ plaquette. In
contrast to the two-center spin-singlet excitation, such a
transition is not accompanied by strong two-magnon Raman
processes, what could be used to identify this kind of exciton.
The redistribution of spin density from copper to oxygen for the
$b_{1g}\rightarrow e_u$ transition switches on a strong
ferromagnetic Heisenberg exchange between nearest neighbor CuO$_4$
plaquettes. The different sign of exchange coupling for the $e_u$
and the $b_{1g}$ holes (ferromagnetic for the former, and
antiferromagnetic for the latter) leads to a number of temperature
anomalies near the antiferromagnetic phase transition. First, one
expects a line broadening and a $blue$ shift for the energy of the
one-center exciton $b_{1g} \rightarrow e_u$ by lowering the
temperature near and below $T_N$. All these expectations are
experimentally found for the $2.0$ eV line in Sr$_2$CuO$_2$Cl$_2$
\cite{Choi} confirming its one-center CT excitonic nature.

\begin{figure}[t]
\includegraphics[width=8cm,angle=0]{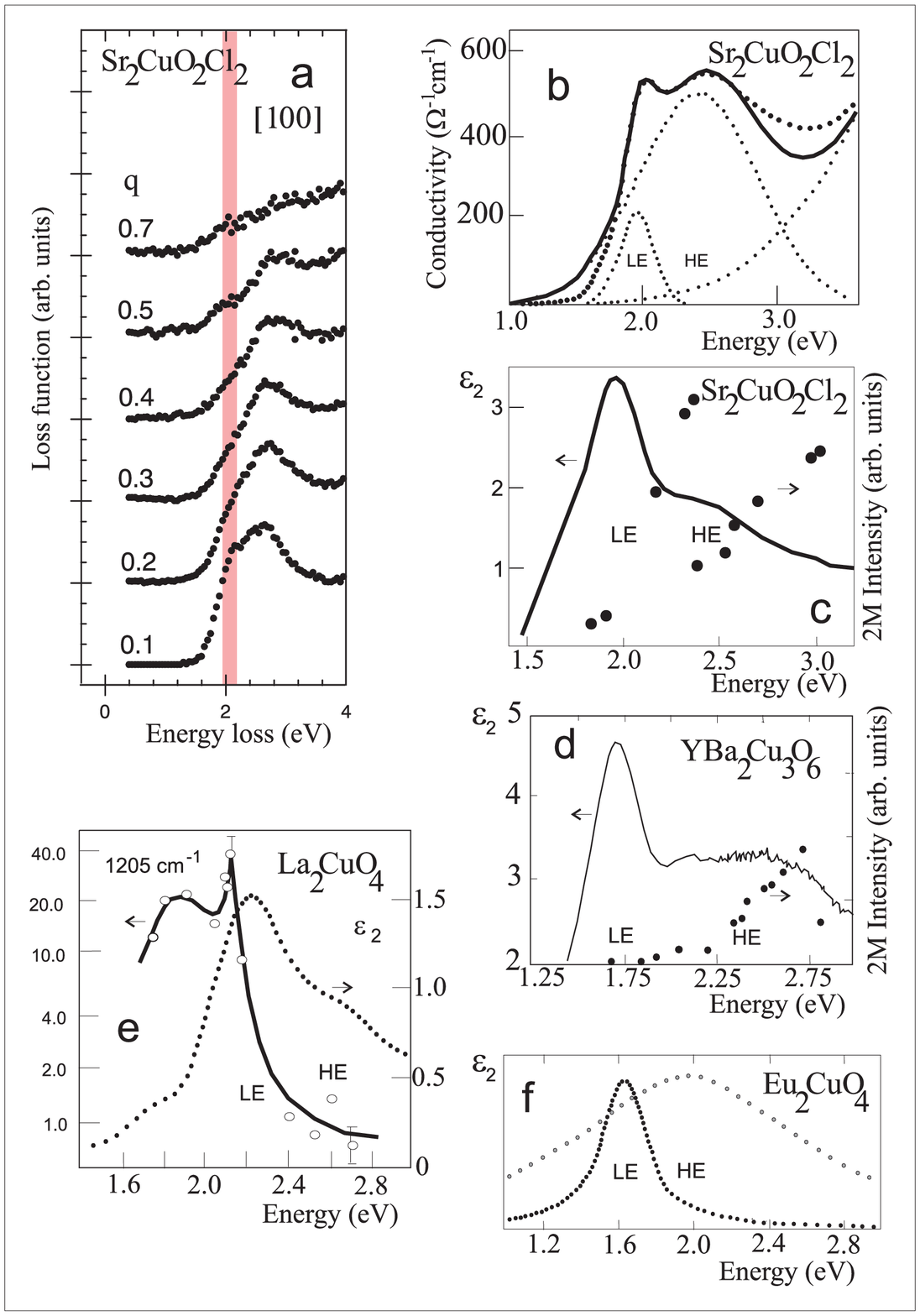}
\caption{Some experimental results of   EELS and optical
measurements for different insulating cuprates which clearly
demonstrate a structure of the optical gap: a) low-energy part of
EELS spectra in Sr$_2$CuO$_2$Cl$_2$ for ${\bf k}\parallel [100]$
\cite{Neudert}; b) optical conductivity in Sr$_2$CuO$_2$Cl$_2$
\cite{Choi};
  c), d) imaginary part of the dielectric response in Sr$_2$CuO$_2$Cl$_2$
 \cite{optics,Blumberg} and YBa$_2$Cu$_3$O$_6$,\cite{Cooper} respectively,
together with
 the integrated intensity of the related two-magnon (2M) Raman spectrum as a
function of the
 laser excitation energy; e) imaginary part of the
 dielectric response in La$_2$CuO$_4$ \cite{Falck} and the normalized
integrated
 intensity of 1205  cm$^{-1}$ Raman line in La$_2$CuO$_4$ as a function of
the
 laser excitation energy;\cite{Ohana} f) contribution of two fitted oscillators
to the imaginary
 part of the dielectric function for the insulating Eu$_2$CuO$_4$.\cite{Krich}}
 \label{fig10}
 \end{figure}

The behavior of one- and two-center excitons in monoxide CuO
\cite{Moskvin1994,Sukhorukov} has some specific features due to the
low (monoclinic) crystal symmetry. The $e_{u}(\pi)$ state is
splitted into two components with markedly differing inter-chain
$e_{u}(\pi) - b_{1g}$ exchange, and the respective CT bands near 1.7
eV  show markedly different temperature behaviors. \cite{Sukhorukov}
Some experimental data  which demonstrate the two-peak  structure of
the CT gap in insulating cuprates are presented in Fig.\
\ref{fig10}.

\section{Conclusion}
We have developed a semi-quantitative cluster approach based on the
complete Cu $3d$-O $2p$ set of the embedded CuO$_4$ cluster orbitals
with a reasonable choice of single-particle and correlation
parameters  to consistently describe the electron-hole excitations
in insulating cuprates in a rather wide energy range up to $10\div
15$ eV.
We extended the Zhang-Ng model by considering the complete set of Cu
$3d$ and O $2p$ orbitals and by introducing  one-center and
two-center excitons. In this connection, it is worth noting that the
small charge transfer exciton of the ZN-model can actually be
attributed namely to  one-center excitons rather than to two-center
ones (see e.g., Fig. 3 in Ref.\ \onlinecite{Ng}). Instead of one
$d-p_{\sigma}$ CT transition of the ZN-model with an energy $\approx
2.5$ eV ($\Gamma$-point) we arrive at a set of one- and two-center
excitons generated by $d-p_{\sigma}$ charge transfer, and occupying
a very broad energy range from $\approx 2$ up to $\approx 13$ eV.
Moreover, the largest dipole intensities have the excitons with the
highest energy  $9.0\div 13.0$ eV.

The simple cluster model allows to consistently account for
correlation effects in the final two-hole states for two-center
excitons. It is shown that these effects are of particular
importance both for the energies and intensities of such excitons.

We argue of an important role played by the transition matrix
element effects both in optical and EELS spectra. In the latter case
we obtain the momentum dependence of matrix elements for different
both intra-center and inter-center transitions.  One should note
that the optical and EELS measurements provide a very important
instrument to inspect the electronic structure and the energy
spectrum of insulating cuprates. However, decisive conclusions might
be taken only after a detailed account for "matrix element" or
intensity effects.

The exciton-exciton interaction is shown to be of particular
importance for the actual momentum dependence of the EELS
intensities being a probable origin of such peculiarities as an
intensity compensation point due to a destructive exciton-exciton
interference.

Starting from the predictions of the model theory and some ARPES
data we were able to give a semi-quantitative description of the
experimental EELS spectra for  Sr$_2$CuO$_2$Cl$_2$ that is free from
many shortcomings of the simplified ZN-model. Both, the different
experimental data and the theoretical analysis show that the nature
of the CT gap in insulating parent cuprates is determined by two
nearly degenerate excitations, a localized intra-center and a
inter-center CT exciton with considerable dispersion. The former
represents the optical counterpart of the so-called "1 eV peak"
revealed by ARPES measurements in a number of cuprates and is
associated with the hole CT transition $b_{1g}\rightarrow
e_{u}(\pi)$ from the Cu $3d$-O $2p$ hybrid $b_{1g}\propto
d_{x^2-y^2}$ state to the purely oxygen O $2p_\pi$ state localized
on the CuO$_4$ plaquette, while the latter corresponds to the
$b_{1g}^b \rightarrow b_{1g}$ CT transition between neighboring
plaquettes with the ZR singlet being the final two-hole state.

Making use of the EELS measurements for the 1D cuprate Sr$_2$CuO$_3$
with two different directions of the transferred momentum, we
demonstrate straightforwardly the two-peak nature of the CT gap with
the coexistence of nearly degenerate one- and two-center CT
excitons. In contrast to the 2D case, the two different types of
excitons may be well separated by choosing the transferred momentum
to be perpendicular to the chain direction.

The structure of the optical gap with two rather well-defined CT
excitons seems to be typical for a wide group of parent cuprates
that implies  a revisit of some generally accepted views on the
electronic structure both of 1D and 2D cuprates.

The nonbonding O $2p_{\pi}$ states with $e_u$ and $a_{2g}$ symmetry
form the energetically  lowest purely oxygen hole states localized
on a CuO$_4$ plaquette with energy $\approx 2$ eV. One might
speculate that these O $2p_\pi$ states could be as preferable for
the localization of additive hole as  the $b_{1g}\propto
d_{x^2-y^2}$ ground state which would result in the instability of
the Zhang-Rice singlet in doped cuprates. The most probable
candidate states for a competition in energy with the Zhang-Rice
$(b_{1g}^{2})$ (or ${}^{1}A_{1g}$) singlet are the singlet
${}^{1}E_{u}$, or the triplet ${}^{3}E_{u}$ terms of the $b_{1g}e_u$
configuration. Such a ${}^{1}A_{1g}-{}^{1,3}E_{u}$ competition for
the hole-doped CuO$_4$ plaquette can result in an unconventional
behavior of doped cuprates. Perhaps, this could substantially
influence the nature of the superconducting state in these systems.
Such a conclusion  is supported both by local-density-functional
calculations, \cite{McMahan1,Mattheiss} {\it ab initio} unrestricted
Hartree-Fock self-consistent field MO method (UHF-SCF) for
copper-oxygen clusters, \cite{Tanaka,Tanaka1} and a large variety of
experimental data. To the best of our knowledge, one of the first
quantitative conclusions on a competitive role of the hybrid
copper-oxygen $b_{1g}(d_{x^2 -y^2})$ orbital and purely oxygen O
$2p_{\pi}$ orbitals in the formation of valence states near the
Fermi level in the CuO$_2$ planes has been made by A.K. McMahan {\it
et al.} \cite{McMahan1} and J. Tanaka {\it et al.} \cite{Tanaka}
Namely these orbitals, as they state, define the low-energy physics
of copper oxides.

Insulating  2D cuprates like Sr$_2$CuO$_2$Cl$_2$  are not molecular
crystals, and making use of the small exciton approach inevitably
leads to problems with strong excitonic overlap and mixing, and
manifestation of  band-like effects owing to electron-hole pair
decoupling. Nevertheless, we suppose that the strong
electron-lattice polarization effects \cite{Overhauser} for CT
states may provide an effective localization  of CT excitons. So,
the small exciton model turns out to be a good and instructive
approximation to describe optical and EELS spectra in different
insulating cuprates with CuO$_4$ plaquettes. In this connection, we
would like to emphasize the specific role played by the ground state
of the cuprate in the formation of optical and EELS response. For a
strongly correlated ground state the main part of the optical and
EELS response intensity is determined by the excitonic sector, while
for a band-like weakly correlated ground state it is the sector of
unbound electron-hole pairs. Moreover, in the case of the
strongly-correlated insulating cuprates it seems misleading to
search for separate contributions of excitons and electron-hole
continuum. \cite{Ng}

Concluding, it should be noted that the quantum-chemical CuO$_4$
cluster model represents a physically  clear albeit rather
simplified approach to consider electron-hole excitations. However,
it seems that such an approach allows to catch the essential physics
of the charge transfer transitions, and should be a necessary step
both in the qualitative and semi-quantitative description of
insulating cuprates.

\section{Acknowledgments}

A.S.M. acknowledges the stimulating discussions with B.B.
Krichevtsov, R.V. Pisarev, N.N. Loshkareva, Yu.P. Sukhorukov, and
support by  SMWK Grant,  INTAS Grant No. 01-0654, CRDF Grant No.
REC-005, RFBR Grant No. 04-02-96077. S.-L.D. and J.M. acknowledge
the support by DFG.


\begin{thebibliography}{99}

\bibitem{ZSA} T.  Zaanen, G.A.  Sawatzky, and J.W.  Allen, Phys.  Rev.  Lett.  {\bf
55}, 418 (1985).

\bibitem{Tokura}
M. Imada, A. Fujimori, and Y. Tokura, Rev. Mod. Phys. {\bf 70}, 1039
(1998).

\bibitem{Moskvin1994}
A.S. Moskvin, N.N. Loshkareva, Yu.P. Sukhorukov, M.A. Sidorov, and
A.A. Samokhvalov, Zh. Eksp. Teor. Fiz. {\bf 105}, 967 (1994) [JETP,
{\bf 78}, 518 (1994)].

\bibitem{Sukhorukov}
Yu.P. Sukhorukov, N.N. Loshkareva, A.A. Samokhvalov, and A.S.
Moskvin, Zh. Eksp. Teor. Fiz. {\bf 108}, 1821 (1995)[JETP, {\bf 81},
998 (1995)].

\bibitem{Falck}
J.P. Falck, A. Levy, M.A. Kastner, and R.J. Birgeneau,  Phys. Rev.
Lett. {\bf 69}, 1109 (1992).

\bibitem{Ohana}
I. Ohana, D. Heiman, M.S. Dresselhaus, and P.J. Picone, Phys. Rev. B
{\bf 40}, 2225 (1989).

\bibitem{Kishida1}
H. Kishida, M. Ono, A. Sawa, M. Kawasaki, Y. Tokura, and H. Okamoto
Phys. Rev. B {\bf 68}, 075101 (2003).

\bibitem{THG}
A. Sch\"{u}lzgen, Y. Kawabe, E. Hanamura, A. Yamanaka, P.-A.
Blanche, J. Lee, H. Sato, M. Naito, N.T. Dan, S. Uchida, Y. Tanabe,
and N. Peyghambarian, Phys. Rev. Lett. {\bf 86}, 3164 (2001).

\bibitem{Krich}
B.B.  Krichevtsov, R.V. Pisarev, A. Burau, H.-J. Weber, S.N. Barilo,
and D.I. Zhigunov, J.Phys.: Cond.Matt. {\bf 6},4795 (1994).

\bibitem{Heyen}
E.T. Heyen, J. Kircher, and M. Cardona,  Phys. Rev. B {\bf 45}, 3037
(1992-II).

\bibitem{Cooper}
S.L. Cooper, D. Reznik, and A. Kotz {\it et al.}, Phys. Rev. B {\bf
47}, 8233 (1993).

\bibitem{Choi}
H.S. Choi, Y.S. Lee, T.W. Noh, E.J. Choi, Yunkyu Bang, and Y.J. Kim,
Phys. Rev. B {\bf  60}, 4646 (1999).

\bibitem{optics}
A. Zibold, H.L. Liu, S.W. Moore, J.M. Graybeal, and D.B. Tanner,
Phys. Rev. B {\bf 53}, 11734 (1996).

\bibitem{Blumberg}
G. Blumberg, P. Abbamonte, M. V. Klein, W.C. Lee,  D.M. Ginsberg,
L.L. Miller, and A. Zibold, Phys. Rev. B {\bf  53}, R11930 (1996).

\bibitem{Schumacher}
A.B. Schumacher, J.S. Dodge, M.A. Carnahan, R.A. Kaindl,  D.S.
Chemla, and L.L. Miller, Phys. Rev. Lett. {\bf 87}, 127006 (2001).

\bibitem{Ogasawara}
T. Ogasawara, M. Ashida, N. Motoyama, H. Eisaki, S. Uchida, Y.
Tokura, H. Ghosh, A. Shukla, S. Mazumdar, and M. Kuwata-Gonokami,
Phys. Rev. Lett. {\bf 85}, 2204 (2000).

\bibitem{nonlinear}
Y. Mizuno, K. Tsutsui, T. Tohyama, and S. Maekawa, Phys. Rev. B {\bf
62}, R4769 (2000-II).

\bibitem{Kishida}
H. Kishida, H. Matsuzaki, H. Okamoto,  T. Manabe, M. Yamashita, Y.
Taguchi, and Y. Tokura, Nature {\bf 405}, 929 (2000).

\bibitem{Wang}
Y.Y. Wang, F.C. Zhang, V.P. Dravid, K.K. Ng, M.V. Klein, S.E.
Schnatterly, and L.L. Miller, Phys. Rev. Lett. {\bf 77}, 1809
(1996).

\bibitem{EELS}
R. Neudert, T. Boeske, O. Knauff, M. Knupfer, M.S. Golden, G.
Krabbes, J. Fink, H. Eisaki, and S. Uchida,  Physica  B {\bf
230-232}, 847 (1997).

\bibitem{EELS1}
J. Fink, R. Neudert, H.C. Schmelz, T. Boeske, O. Knauff, S. Haffner,
M. Knupfer, M.S. Golden, G. Krabbes,  H. Eisaki, and S. Uchida,
Physica B {\bf 237-238},  93 (1997).


\bibitem{Abbamonte}
P. Abbamonte, C.A. Burns, E.D. Issacs, P.M. Platzman, L.L. Miller,
S.W. Cheong, and M.V. Klein, Phys. Rev. Lett. {\bf 83}, 860 (1999).

\bibitem{Hasan}
M.Z. Hasan, E.D. Issacs, Z.-.X. Shen,  L.L. Miller, K. Tsutsui, T.
Tohyama, and S. Maekawa, Science, {\bf 288}, 1811 (2000); M.Z.
Hasan, E.D. Issacs, Z-.X. Shen,  and L.L. Miller, cond-mat/0102492.


\bibitem{ZR}
F.C. Zhang and  T.M. Rice,  Phys. Rev. B {\bf 37}, 3759 (1988).

\bibitem{Wagner}
J. Wagner, W. Hanke, and D.J. Scalapino, Phys. Rev. B {\bf 43},
10517 (1991).

\bibitem{Guo}
D. Guo, S. Mazumdar, S.N. Dixit, F. Kajzar,  F. Jarka, Y. Kawabe,
and N. Peyghambarian, Phys. Rev. B {\bf  48}, 1433 (1993-I).

\bibitem{Simon}
M.E. Simon, A.A. Aligia, C.D. Batista, E.R. Gagliano, and F. Lema,
Phys. Rev. B  {\bf 54}, R3780 (1996-II); M.E. Simon, A.A. Aligia,
and E.R. Cagliano, cond-mat/9707128.

\bibitem{Ng} F.C. Zhang and K.K. Ng, Phys. Rev. B {\bf 58}, 13520 (1998).

\bibitem{Hanamura}
Eiichi Hanamura, Nguen Trung Dan, and Yukito Tanabe,  Phys. Rev. B
{\bf 62}, 7033 (2000).

\bibitem{Kuzian}
R.O. Kuzian, R. Hayn, and A.F. Barabanov, Phys. Rev. B {\bf 68},
195106 (2003).

\bibitem{PRB}
A.S. Moskvin, R. Neudert, M. Knupfer, J. Fink, and R. Hayn, Phys.
Rev. B {\bf 65}, 180512(R) (2002).

\bibitem{PRL}
A. S. Moskvin, J. M\'{a}lek, M. Knupfer, R. Neudert, J. Fink, R.
Hayn, S.-L. Drechsler, N. Motoyama, H. Eisaki, and S. Uchida, Phys.
Rev. Lett. {\bf 91}, 037001 (2003).


\bibitem{Penc}
W. Stephan and K. Penc, Phys. Rev. B {\bf 54}, R17 269 (1996); K.
Penc and W. Stephan, Phys. Rev. B {\bf 62}, 12 707 (2000).

\bibitem{Tsutsui}
K. Tsutsui, T. Tohyama, and S. Maekawa, Phys. Rev. B {\bf 61}, 7180
(2000).

\bibitem{Kim04}
Y.-J. Kim, J.P. Hill, H. Benthien, F.H.L. Essler, E. Jeckelmann,
H.S. Choi, T.W. Noh, N. Motoyama, K.M. Kojima, S. Uchida, D. Casa,
and T. Cog, Phys. Rev. Lett. {\bf 92}, 137 402 (2004).


\bibitem{Eskes}
H. Eskes,  L.H. Tjeng, and  G.A. Sawatzky,  Phys. Rev. B {\bf 41},
288 (1990).

\bibitem{Ghijsen}
J. Ghijsen, L.H. Tjeng, J. van Elp, H. Eskes, J. Westerink,  G.A.
Sawatzky, and M.T. Czyzyk, Phys. Rev. B {\bf  38}, 11322 (1988).


\bibitem{Stechel}
E.B. Stechel and D.R. Jennison,  Phys. Rev. B  {\bf 38}, 8873
(1988).

\bibitem{McM}
A.K. McMahan,  J.F. Annett, and R.M. Martin, Phys. Rev. B {\bf 42},
6268 (1990).

\bibitem{Czyzyk}
M.T. Czyzyk and  G.A. Sawatzky,   Phys. Rev. B {\bf 49}, 14211
(1994-II).

\bibitem{Mattheiss}
L.F.\ Mattheiss and D.R. Hamann, Phys.\ Rev.\ B {\bf 40}, 2217
(1989).

\bibitem{Hayn}
R. Hayn, H. Rosner, V. Yu. Yushankhai, S. Haffner, C. Duerr, M.
Knupfer, G. Krabbes, M. S. Golden, J. Fink, H. Eschrig, D. J. Singh,
N.T. Hien, A.A. Menovsky, Ch. Jung, and G. Reichardt, Phys. Rev. B
{\bf  60}, 645 (1999).

\bibitem{Pothuizen}
 J.J.M. Pothuizen, R. Eder, N.T. Hien,  M. Matoba, A.A. Menovsky, and G.A.
Sawatzky, Phys. Rev. Lett. {\bf 78}, 717 (1997).

\bibitem{Duerr}
C. Duerr, S. Legner, R. Hayn, S.V. Borisenko, Z. Hu, A. Theresiak,
M. Knupfer, M. S. Golden, J. Fink, F. Ronning, Z.-X. Shen, H.
Eisaki, S. Uchida, C. Janowitz, R. Mueller, R.L. Johnson, K.
Rossnagel, L. Kipp, and G. Reichardt, Phys. Rev. B {\bf 63},
014505-1 (2000).

\bibitem{Overhauser}
Albert W. Overhauser,  Phys. Rev. {\bf 101}, 1702 (1956).

\bibitem{Boman}
M. Boman and R. J. Bursill, Phys. Rev. B  {\bf 57}, 15167 (1998).

\bibitem{Cherepanov}
V.I. Cherepanov, E.N. Kondrashov and A.S. Moskvin, Sov.Phys.-Solid
State, {\bf 42}, 866  (2000).

\bibitem{bersuker}
I.B. Bersuker and V.Z. Polinger, Vibronic Interactions in Molecules
and Crystals, Springer-Verlag, Berlin, 1989.

\bibitem{Hybertsen}
M.S. Hybertsen, M. Schluter, and N.E. Christensen, Phys. Rev. B {\bf
39},  9028 (1989).

\bibitem{Hybertsen1}
M.S. Hybertsen, E.B. Stechel, M. Schluter, and D.R. Jennison, Phys.
Rev. B {\bf 41}, 11068 (1990).

\bibitem{Roland}
V.Yu. Yushankhai, V.S. Oudovenko, and R. Hayn, Phys. Rev. B {\bf
55}, 15562 (1997-I).

\bibitem{Davydov}
A.S. Davydov, Theory of Molecular Excitons, McGraw-Hill, New York,
1962.

\bibitem{CuBO}
M. Martinez-Ripoli, S. Martinez-Carrera, and S. Garsia-Blanco, Acta
Crystallogr. Sect. B {\bf 27}, 677 (1971).

\bibitem{x-ray}
T. Boeske, O. Knauff, R. Neudert {\it et. al.}, Phys. Rev. B {\bf
56}, 3438 (1997).

\bibitem{Fujimori}
A. Fujimori, Y. Tokura, H. Eisaki, H. Takagi, S. Uchida, and M.
Sato, Phys. Rev. Lett. {\bf 40}, (1989) 7303.

\bibitem{XAS}
S. Haffner, R. Neudert, M. Kielwein {\it et. al.},  Phys. Rev. B
{\bf B 57}, 3672 (1998).

\bibitem{XAS1}
T. Boeske, K. Maiti, O. Knauff {\it et. al.},  Phys. Rev. B {\bf
57}, 138 (1997).

\bibitem{LaRosa}
S. LaRosa, I. Vobornik, F. Zwick {\it et.al.}, Phys. Rev. B {\bf
56}, R525 (1997).

\bibitem{Kim}
 C. Kim, P.J. White, Z.-X. Shen {\it et. al.},  Phys. Rev. Lett. {\bf
80}, 4245 (1998).

\bibitem{Ronning}
F.~Ronning, C.~Kim, D.L.~Feng {\it et. al.}, Science {\bf 282}, 2067 (1998).

\bibitem{Haffner1}
S. Haffner, D. M. Brammeier, C. G. Olson {\it et. al.}, Phys. Rev. B
{\bf  61}, 14 378 (2000).

\bibitem{Wells}
B.O. Wells, Z.-X. Shen, A. Matsuura, D. M. King, M. A. Kastner, M.
Greven, and R. J. Birgeneau, Phys. Rev. Lett.  {\bf 74}, 964 (1995).

\bibitem{Atzkern}
S. Atzkern, M. Knupfer, M.S. Golden {\it et. al.},  Phys. Rev. B
{\bf B 62}, 7845 (2000-II).

\bibitem{Li}
Y. Mizuno, T. Tohyama, S. Maekawa {\it et. al.},  Phys. Rev. B {\bf
57}, 5326 (1998-I).

\bibitem{Drechsler}
R. Neudert, S.-L. Drechsler, J. Malek {\it et. al.},  Phys. Rev. B
{\bf 62}, 10752 (2000-I).

\bibitem{Sturge}
{\it  Excitons},  E.I. Rashba, M.D. Sturge eds., NH, 1987.

\bibitem{Loudon}
P.A. Fleury and R. Loudon,  Phys. Rev. {\bf 166}, 514 (1968).

\bibitem{Pisarev}
R.V. Pisarev, I. S\"{a}nger, G. A. Petrakovskii, and M. Fiebig,
Phys. Rev. Lett. {\bf 93}, 037204 (2004).

\bibitem{Bassi}
M. Bassi, P. Camagni, R. Rolli, G. Samoggia, F. Parmigiani, G.
Dhalenne, and A. Revcolevschi, Phys. Rev. B {\bf 54}, R11030
(1996-II).

\bibitem{Knupfer}
M. Knupfer, private communication.

\bibitem{Neudert}
R. Neudert, Ph.D. Thesis, University of Technology Dresden, 1999.

\bibitem{1D-EELS}
R. Neudert, M. Knupfer, M.S. Golden, J. Fink, W. Stephan, K. Penc,
N. Motoyama, H. Eisaki, and S. Uchida,  Phys. Rev. Lett. {\bf 81},
657 (1998).

\bibitem{McMahan1}
A.K. McMahan, R.M. Martin, and S. Satpathy, Phys. Rev. B {\bf 38},
6650 (1988).

\bibitem{Tanaka}
Jiro Tanaka, Koji Kamiya, and Chizuko Tanaka, Physica C {\bf 61},
451 (1989).

\bibitem{Tanaka1}
J. Tanaka and C. Tanaka, J. Phys. Chem. Solids {\bf 59}, 1861
(1998).

\bibitem{Hill}
J.P. Hill, C.-C. Kao, W.A.L. Caliebe,  M. Matsubara, A. Kotani, J.L.
Peng, and R.L. Greene,  Phys. Rev. Lett. {\bf 80}, 4967 (1998).


\end{thebibliography}
\end{document}